\documentclass[superscriptaddress,twocolumn,aps,pre,showpacs]{revtex4}
\usepackage{epsfig}
\usepackage{graphicx}
\usepackage{amssymb,amsfonts,amsmath}
\usepackage{dcolumn}
\usepackage{bm}
\begin{document}

\title{Solvable random walk model with memory and its relations 
with Markovian models of anomalous diffusion}

\author{D. Boyer}
\email{boyer@fisica.unam.mx}
\affiliation{Instituto de F\'\i sica, Universidad Nacional Aut\'onoma de 
M\'exico, D.F. 04510, M\'exico}
\affiliation{Centro de Ciencias de la Complejidad, Universidad Nacional 
Aut\'onoma de M\'exico, D.F. 04510, M\'exico}

\author{J. C. R. Romo-Cruz}
\affiliation{Instituto de F\'\i sica, Universidad Nacional Aut\'onoma de 
M\'exico, D.F. 04510, M\'exico}

\date{\today}

\begin{abstract}
Motivated by studies on the recurrent properties of animal 
and human mobility, we introduce a path-dependent random walk model 
with long range memory for which not only the mean square displacement (MSD)
can be obtained exactly in the asymptotic limit, but also the
propagator. The model consists of a random walker 
on a lattice, which, at a constant rate, stochastically relocates at
a site occupied at some earlier time. This time in the past 
is chosen randomly according to a memory kernel, whose temporal
decay can be varied via an exponent parameter. In the weakly 
non-Markovian regime, memory reduces the diffusion coefficient from the 
bare value. When the mean backward jump in time diverges, the diffusion
coefficient vanishes and a transition to an anomalous subdiffusive 
regime occurs. Paradoxically, at the transition, the process 
is an anti-correlated L\'evy flight. Although in the subdiffusive regime 
the model exhibits some features of the continuous time random walk 
with infinite mean waiting time, it belongs to another universality class.
If memory is very long-ranged, a second transition 
takes place to a regime characterized by a logarithmic growth of the MSD 
with time. In this case the process is asymptotically Gaussian and effectively 
described as a scaled Brownian motion with a diffusion coefficient 
decaying as $1/t$. 
\end{abstract}

 \maketitle

\section{Introduction}

\lq\lq True" self-interacting random walks (as opposed to static 
models like the self-avoiding walk of polymer physics) 
are an important class of non-Markovian kinetic
processes with long range memory \cite{annals}. 
Reinforced random walks are self-attracting processes where, 
typically, a random walker tends to preferentially revisit the 
nearest-neighbor sites that it has visited before (or, in a variant, 
recross the edges already crossed).
In the last few decades, 
reinforced random walks have received a particular attention, not only for
the mathematical challenges they raise, but also for their 
applications to biology \cite{siam,pemantle}. These walks have been used for
the description of the displacements of ants or bacteria \cite{siam}. 
In ecology, they can also
represent simple models of \lq\lq site fidelity", a behavior observed
in many animals in the wild \cite{gautestad2005,gautestad2006}. 
Many reinforced walk models are defined through
transition probabilities that depend on the number of 
visits (or crossings) received by the sites (or edges) and the
resulting dynamics is thus strongly path-dependent.

Some rigorous results have been obtained for certain reinforcement rules, 
showing that different dynamical behaviors can emerge depending on the 
strength of memory and the spatial dimension. A single walker may 
asymptotically become localized (keeping oscillating between a few sites), 
or diffusive, in which the range of its position $X_t$ is infinite and 
the origin visited infinitely often in $1d$ \cite{davis,annals,siam}. 
Numerical simulations actually show that a variety of models seems to 
exhibit a 
phase transition at finite reinforcement between a localized and 
a diffusive regime \cite{havlin,woo,grassberger,epl}. Interestingly,
in the diffusive regime, the same studies have presented
evidence that diffusion is anomalous, namely, subdiffusive. This type of
motion, widely studied in Markovian contexts \cite{bouchaud,klafter}, is 
characterized by a mean 
square displacement (MSD) of the particle which does not follow the
Smoluchowski-Einstein law of Brownian motion, but a slower one, of the form
$\langle X_t^2\rangle\propto t^{\mu}$ with $\mu$ an exponent $<1$.

Few random walk models with long range memory are
analytically tractable \cite{elephant1,elephant2,italian}.
For a single particle,
path-dependent diffusion is notoriously difficult to formalize
and many results on reinforced random walks are based on numerical
simulations. Usually,
such processes cannot be described by a master equation
for the single-time occupation probability (or propagator).
Instead, they require the introduction of multiple-time distribution 
functions, that are related to each other via a hierarchy of relations.
Path integral approaches \cite{peliti,nuovo} or approximate scaling 
arguments \cite{quasistatic,song} have been developed to overcome these 
difficulties. Nonlinear integro-differential Fokker-Planck 
equations can also be derived within a mean-field approximation \cite{epl}.
This latter approach can give a fairly good picture of the phase 
diagram but remains limited for a precise dynamical description.
 
The recent ecological literature reveals a regain of interest
for reinforced random walks. Thanks to spectacular 
advances in tracking technology, the positions 
of individual
animals \cite{nathan,sims,weimer}, including 
humans \cite{geisel,gonzalez,song1,song}, can be recorded with a high 
resolution and during long periods of time. Recently, random walk models 
with memory have successfully explained some features of
the trajectories of humans \cite{song},
monkeys \cite{boyersolis} or bisons \cite{morales2014}, 
which all exhibit 
strong recurrence, an anomalous slow diffusion and an heterogeneous 
occupation of space. Other theoretical studies have identified memory 
as a key factor for
the emergence of home ranges, that is, a restricted space use 
\cite{moorcroftlewis,borger,vanmoorter,boyerwalsh,oikos2013}.
Reinforced random walks thus offer a 
promising alternative to Markovian random walks, which 
remain the dominant modeling paradigm in movement ecology 
\cite{turchin,colding,randomsearch}. However, most
models developed so far are computational  
and we still lack a basic understanding of the effects of 
memory on mobility patterns, even in simple cases. 
For this purpose, it is desirable, in parallel with
ecologically realistic computational approaches, 
to gain knowledge from the mathematical analysis of very simple models.

Here, we exactly solve in the asymptotic time limit a reinforced 
model, that generalizes the one presented in \cite{boyersolis}, where a random
walker intermittently relocates to sites visited in the past. We obtain
not only the behavior of the MSD, but also the properties of the propagator 
or diffusion front, a quantity which is almost unknown for single 
reinforced walks. In contrast with usual reinforced models, here the walker's 
steps are not necessarily directed to nearest-neighbor sites. This property 
has two advantages: it agrees with the empirical fact that many animals 
and humans actually perform long,
intermittent commuting bouts to visit places that are beyond their 
perception range \cite{interm1,interm2,interm3,interm4},
and it also greatly simplifies the mathematical analysis since an
exact master equation can be written in this case.

Depending on a memory parameter, we find that motion can be Brownian or 
subdiffusive. Two subdiffusive regimes are identified, with power-law 
and logarithmic dynamics, respectively. Somewhat paradoxically, a 
L\'evy-like distribution for the length of the steps emerges at the onset
of subdiffusion. We thus suggest that a mechanism based on memory
could be at the origin of the
L\'evy flight patterns observed in many animals 
\cite{levy1,levy2,randomsearch,geisel,sims,rhee}.
In addition, the availability of the propagator
allows us to discuss in a formal way some analogies and differences 
between the present model and well-known, essentially Markovian 
models of anomalous diffusion: notably, the continuous time 
random walk (CTRW) of Montroll and Weiss \cite{montroll}, 
and the scaled Brownian motion \cite{effecFP}.

\section{Model and basic quantities}

Let us consider a one-dimensional lattice with unit spacing and a
walker with position $X_t$ at time $t$. Time is discrete ($t=0,1,2...$)
and the walker starts at the origin, $X_0=0$. Let $q$ be a constant
parameter, $0<q<1$. 
At each time step, $t\rightarrow t+1$, the walker chooses one of the
two movement modes:

{\it (i)} with probability $1-q$, the walker
performs a random step to a nearest-neighbor site, like in the 
simple symmetric random walk;

{\it (ii)} with the complementary probability $q$, the walker 
chooses a random integer $t'$ in the interval $[0,t]$ according to a
probability distribution $p_t(t')$, which is given {\it a priori}. 
Then, the walker directly relocates at the site that it occupied at time $t'$, 
{\it i.e.}, $X_{t+1}=X_{t'}$.

\begin{figure}
\begin{center}
\epsfig{figure=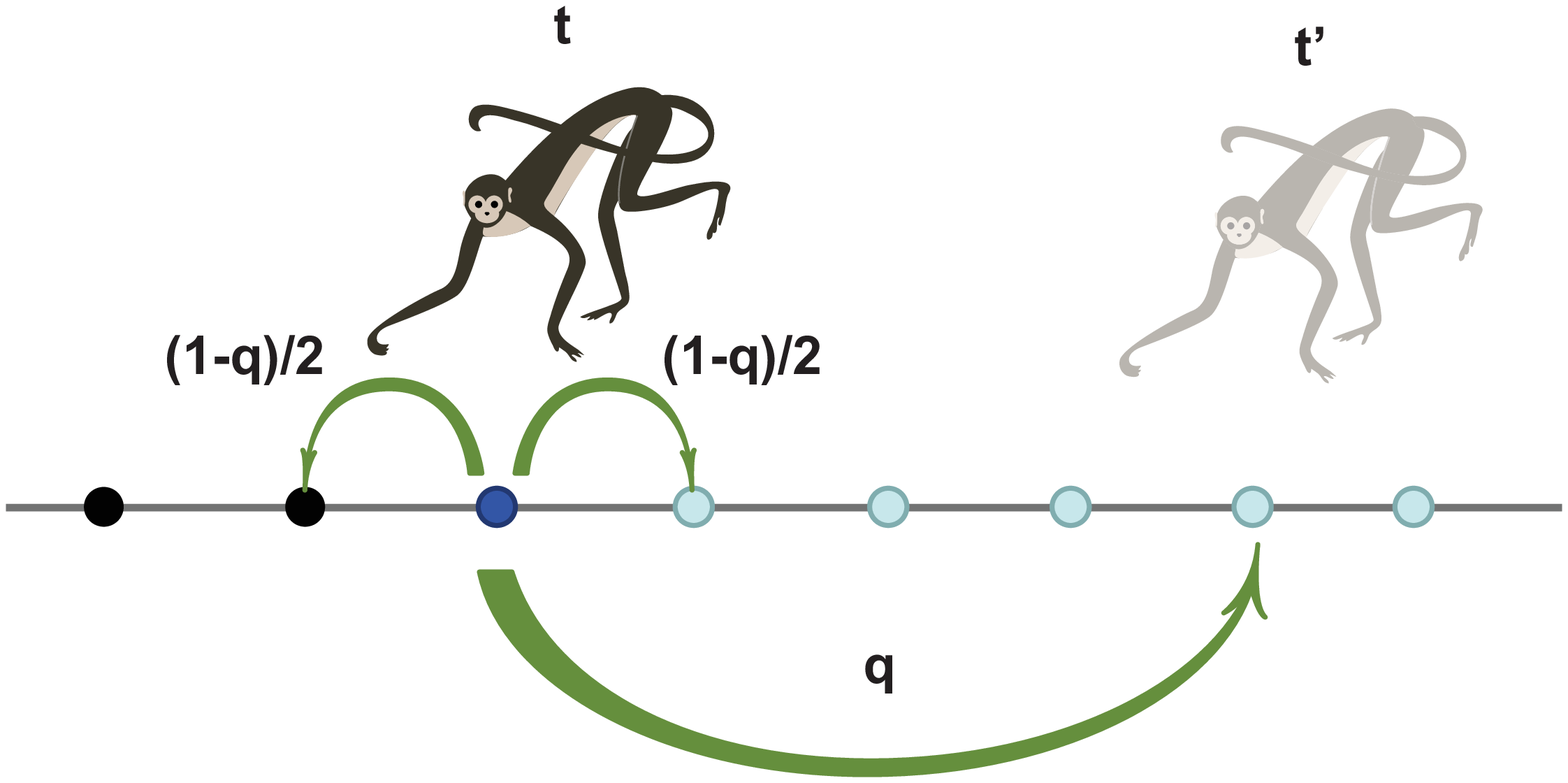,width=3.in}
\epsfig{figure=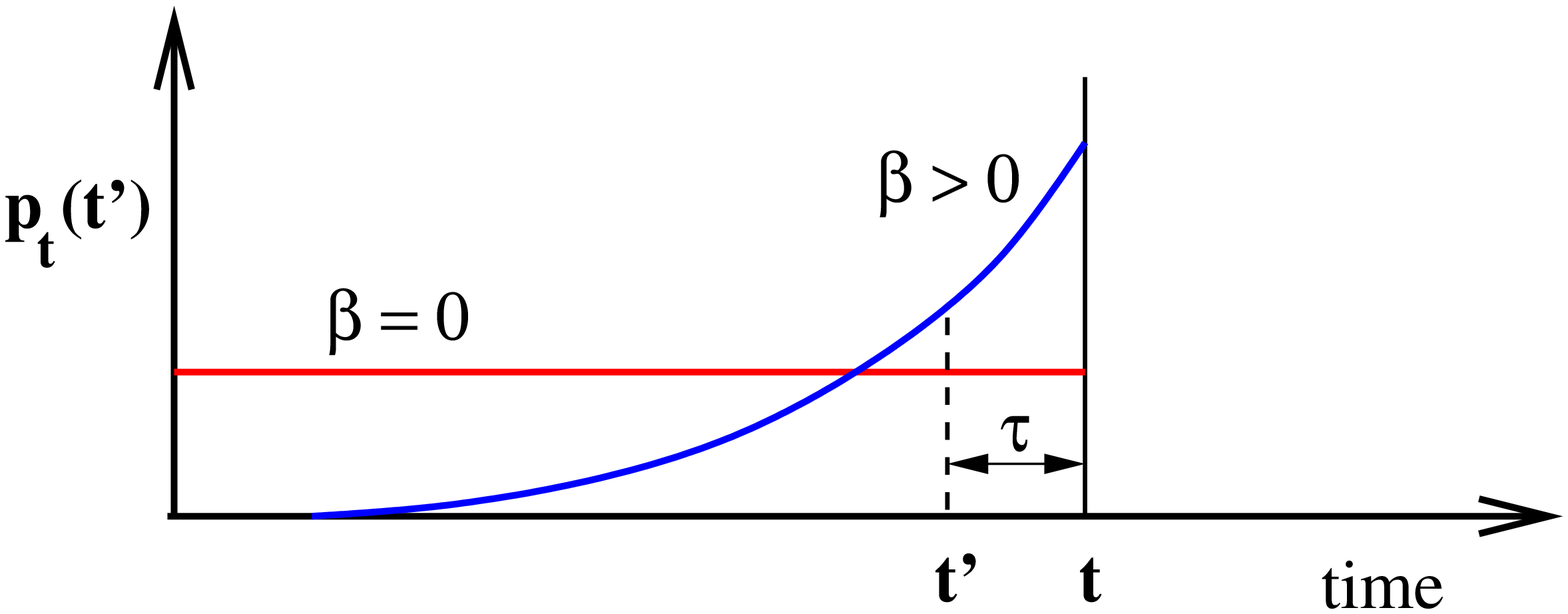,width=3.in}
\end{center}
\vspace{-0.5cm}
\caption{(Color online) Top: At time $t$, the model walker performs a 
nearest-neighbor random walk step with probability $1-q$, or relocates, 
with probability $q$,
to the site it occupied at some earlier time $t'$. Bottom: probability 
distribution of $t'$ for the uniform preferential visit case [$\beta=0$
in Eq. (\ref{Fttp})] 
or in a case with memory decay ($\beta>0$).}
\label{fig:model}
\end{figure}

These rules are depicted in Figure \ref{fig:model}.
If, for instance, $p_t(t')$ is uniform, {\it i.e.}, $p_t(t')=\frac{1}{t+1}$, 
one recovers the preferential visit model 
studied in \cite{boyersolis} (see also \cite{gautestad2005,gautestad2006}). 
In this case, the memory rule {(\it ii)} is equivalent
to revisiting an already visited site, say $n$, with probability 
proportional to the total amount of time spent by the walker at this site
since $t=0$. Hence, 
the more visits site $n$ receives, more likely it will be chosen for future
visits in the memory movement mode. This is the principle of reinforced random 
walks. In a different context, similar rules characterize network growth 
models with preferential attachment \cite{yule,barabasi,leyvraz}. 
An important difference with usual reinforced random walks 
is that, here, any of the previously visited sites is susceptible to
receive the next visit, and not just the 
nearest-neighbors of the current position of the walker. This property 
considerably simplifies the analysis, which still
is not trivial. Note that a Markovian limiting case
of this model is the random walk with stochastic
reseting to the origin \cite{evansmaj}, which corresponds
to $p_{t}(t')=\delta_{t',0}$. 
 
We focus on the propagator $P(n,t)$ of the single particle, 
which is the probability that $X_t=n$ given that $X_0=0$. As 
formally shown in the Appendix \ref{master},
despite of the fact that our process is highly non-Markovian, 
the evolution of $P(n,t)$ is exactly described through a single
master equation:
\begin{eqnarray}\label{genreset}
P(n,t+1)&=&\frac{1-q}{2}P(n-1,t)+\frac{1-q}{2}P(n+1,t)\nonumber\\
&+&q\sum_{t'=0}^{t}p_{t}(t')P(n,t').\ 
\end{eqnarray}
The last term is the probability to choose site $n$ using the memory
mode. This term can also be interpreted as 
proportional to the weighted number of previous visits received by site
$n$, where the weight of a visit received at time $t'$ is $p_t(t')$.
In this study, we will restrict to the cases where the 
probability distribution of $t'$ is of the form 
$p_t(t')\propto F(t-t')$. Hence, the memory kernel $F(\tau)$ depends on the
time elapsed between the remembered event and the present time, 
$\tau\equiv t-t'$.
The normalization condition $\sum_{t'=0}^tp_t(t')=1$ at each $t$ imposes that:
\begin{equation}
p_t(t')=\frac{F(t-t')}{C(t)},
\end{equation}
where
\begin{equation}
C(t)=\sum_{t'=0}^{t}F(t-t').
\end{equation}
In the uniform preferential visit model, $dF(\tau)/d\tau=0$, {\it i.e.} 
memory does not decay. Here, we are 
interested in cases where $dF(\tau)/d\tau<0$, {\it i.e.}, where
recent visits count more than visits performed further in the past, as
represented in Figure \ref{fig:model}-bottom.
We will consider a particularly interesting case, namely,
scale-free memory decays:
\begin{equation}\label{Fttp}
F(\tau)=(\tau+1)^{-\beta},\quad 0\le \tau\le t,
\end{equation}
with $\beta$ an exponent. If $\beta$ is large, the
early time trajectory ($t'\ll t$, or $\tau$ large) tends to be completely 
forgotten. Hence, when the walker uses its memory,
it will most likely decide to revisit
positions occupied at times proximate to the present time $t$. But if $\beta$
is small enough, memory becomes long range and may drastically affect
the diffusion process, like in the preferential visit model ($\beta=0$)
\cite{boyersolis}.

\begin{figure}
\begin{center}
\epsfig{figure=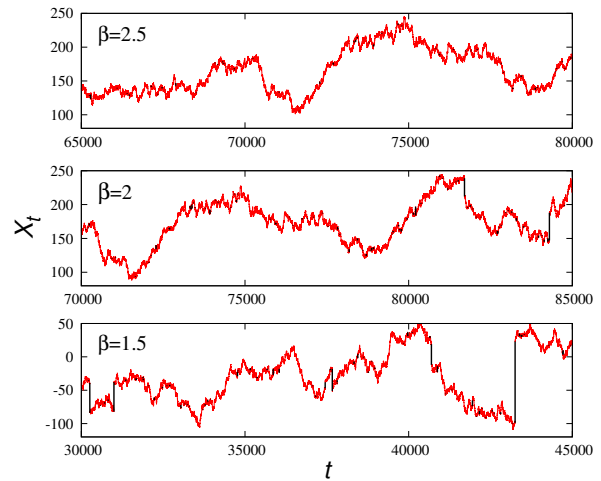,width=2.5in,angle=-90}
\end{center}
\vspace{-0.5cm}
\caption{(color online) Trajectories obtained from numerical simulations 
of the model in
the normally diffusive regime (top), at the subdiffusive threshold
(middle) and in the first subdiffusive regime (bottom). The thick (black) lines
indicate the relocation steps in the memory mode.
The rate of memory use is $q=0.01$ in these cases.}
\label{fig:RelMemTraj}
\end{figure}

At a given time $t$, it is natural to define the mean backward jump in time as
$\langle\tau\rangle_t\equiv\langle t-t'\rangle$, or: 
\begin{equation}\label{deftau}
\langle\tau\rangle_t=\frac{\sum_{\tau=0}^{t}\tau F(\tau)}
{\sum_{\tau=0}^{t}F(\tau)}.
\end{equation}
If $\beta>2$ in the kernel (\ref{Fttp}), 
both sums in Eq. (\ref{deftau}) converge and $\langle\tau\rangle_t$ tends 
to a constant at large $t$:
\begin{equation}
\langle\tau\rangle_{\infty}=\frac{\sum_{\tau=0}^{\infty}\tau(1+\tau)^{-\beta}}
{\zeta(\beta)}
\end{equation}
where $\zeta(\beta)=\sum_{\tau=0}^{\infty}(1+\tau)^{-\beta}$ is the Riemann 
Zeta function. In this case, memory has a finite range and should not affect 
in an essential way the normal diffusion process. Figure \ref{fig:RelMemTraj} 
displays examples of walks generated numerically in $1d$ in three different 
cases: $\beta>2$, $\beta=2$ and $\beta<2$.

\section{Summary of the main results}

The analytical results presented below are derived
in Section \ref{sec:der} using the following methodology.
Instead of solving Eq.(\ref{genreset}), 
we have considered the moments of $P(n,t)$,
defined by the ensemble averages $\langle X_t^{2p}\rangle$. The moments 
obey simpler equations and carry information on the 
behavior of the distribution itself. These equations can be solved in the
asymptotic time limit. We first study the second moment
($p=1$), {\it i.e.} the mean square displacement (MSD). For $\beta>1$, we 
assume that the leading asymptotic term of the MSD is of the form 
$Kt^{\mu}$ and calculate the constants $K$ and $\mu$. For the case 
$\beta<1$, the correct ansatz is $\langle X_t^{2}\rangle\simeq K\ln t$
and we determine $K$.
We next assume that $P(n,t)$ obeys a scaling form at large time in each
case, see Eq.(\ref{scalingP}) below, which enable us to obtain the 
asymptotic behavior of all the higher order moments, for any positive 
integer $p\ge 1$. The MSD ansatzs and the scaling assumption yield consistent 
results, which are also checked numerically by simulations or exact 
time integration of the equations.

\subsection{Mean square displacement $\langle X_t^2\rangle$}

We summarize the exact asymptotic results obtained for 
the MSD, which is the second moment of $P(n,t)$:
\begin{equation}
M_{2}(t)\equiv\sum_{n=-\infty}^{\infty}n^{2}P(n,t)=\langle X_t^2\rangle.
\end{equation}

{\it $\bullet$} If $\beta>2$, we have:
\begin{equation}\label{iv}
M_2(t)\simeq 
\left(\frac{1-q}{1+q\langle\tau\rangle_{\infty}}\right)t.
\end{equation}
This result actually applies to any kernel $F(\tau)$ 
with finite first moment. Thus, diffusion is normal and the main effect 
of memory is to decrease the diffusion constant compared to that of the 
memoryless random walk $(q=0)$. The larger $q$ (frequent memory use) and 
the larger $\langle\tau\rangle_{\infty}$ (better memory), the slower 
diffusion is.

{\it $\bullet$} If $1<\beta<2$, Eq.(\ref{iv}) above no longer holds since 
$\langle\tau\rangle_{\infty}=\infty$, making the 
diffusion constant vanish. Instead,
motion is asymptotically subdiffusive:
\begin{equation}\label{v}
M_2(t)\simeq K t^{\beta-1}\quad{\rm with}\quad 
K=\frac{(1-q)\zeta(\beta)}
{q \int_0^{1}du\frac{1-u^{\beta-1}}
{(1-u)^{\beta}}}.
\end{equation}

{\it $\bullet$} If $\beta<1$, diffusion is even slower, logarithmic in time:
\begin{equation}\label{vi}
M_2(t)\simeq K \ln t\quad{\rm with}\quad 
K=\frac{1-q}{q(1-\beta)\int_0^{1}du\frac{-\ln u}
{(1-u)^{\beta}}}.
\end{equation}
In particular, by setting $\beta=0$ in (\ref{vi}) one recovers
the result of the uniform preferential visit model \cite{boyersolis}, 
$M_2(t)\simeq\frac{1-q}{q}\ln t$.

\subsection{Scaling functions}

In each case above, we make a scaling hypothesis for 
the probability density in the long time limit:
\begin{equation}\label{scalingP}
P(n,t)\simeq\frac{1}{\sqrt{M_2(t)}}\ g\left(\frac{n}{\sqrt{M_2(t)}}\right),
\end{equation}
where, by construction, $g(x)$ is a normalized scaling function 
($\int_{-\infty}^{\infty}g(x)dx=1$)
of unit variance ($\int_{-\infty}^{\infty}x^2g(x)dx=1$). Although $g(x)$
may not be always obtained explicitly, its properties can be inferred from 
the asymptotic behavior 
of the even moments, defined as:
\begin{equation}
M_{2p}(t)=\sum_{n=-\infty}^{\infty}n^{2p}P(n,t),
\end{equation}
$p$ being a positive integer ($M_{2p+1}(t)=0$ by symmetry). 
Assuming that Eq. (\ref{scalingP}) holds, it is easy to see that in 
the long time limit the moments are given by
\begin{equation}\label{scaling2}
M_{2p}(t)\simeq a_p[M_2(t)]^p,
\end{equation}
where $a_p$ is a constant given by 
$a_p=\int_{-\infty}^{\infty}dx\ x^{2p}g(x)$. Obviously, $a_0=a_1=1$.
The constants $a_p$ are calculated in Section \ref{sec:der}.B, and
we find that:

$\bullet$ For $\beta>2$, the scaling function $g(x)$ is Gaussian, {\it i.e.},
\begin{equation}\label{gaussian}
\frac{a_p}{a_{p-1}}=2p-1,
\end{equation}
for any positive integer $p$.

$\bullet$ For $\beta<1$ (logarithmic diffusion case), the process is also
Gaussian asymptotically, namely, $a_p/a_{p-1}=2p-1$ as above, despite of 
the fact that memory is very long ranged and diffusion strongly
anomalous. This result is non-trivial:
the mechanism that makes $g(x)$ Gaussian is strongly driven by memory and
different from the one that leads to Gaussianity in
the Central Limit Theorem (previous case).

$\bullet$  In the intermediate case $1<\beta<2$ (anomalous diffusion 
as $t^{\beta-1}$), $g(x)$ deviates from the Gaussian form. Its moments
satisfy:
\begin{equation}\label{nongaussian}
\frac{a_p}{a_{p-1}}=(2p-1)\frac{p{\cal I}_1(\beta)}
{{\cal I}_p(\beta)}.
\end{equation}
where 
\begin{equation}\label{calI}
{\cal I}_p(\beta)=\int_0^1du\frac{1-u^{p(\beta-1)}}{(1-u)^{\beta}}.
\end{equation}
Note that both (\ref{gaussian}) and (\ref{nongaussian})-(\ref{calI}) 
are {\it universal}
relations, {\it i.e.} the corresponding limiting distributions $P(n,t)$ 
do not depend on $q$ nor on the details of the memory kernel
but only on the fact that $F(\tau)\sim \tau^{-\beta}$ at large $\tau$.

Alternatively, expression (\ref{nongaussian})-(\ref{calI}) can be written 
using special functions. In this paper we define
\begin{equation}
\mu\equiv\beta-1.
\end{equation} 
By integrating 
(\ref{calI}) by parts, we have
${\cal I}_p(\beta)=-\frac{1}{\mu}+pB\left[p\mu,1-\mu\right]$,
with $0<\mu<1$ and
where $B(x,y)$ is the Beta function. Using properties of
the $\Gamma$ function \cite{gamma}, 
Eqs. (\ref{nongaussian})-(\ref{calI}) can be rewritten as:
\begin{equation}\label{momentbis}
\frac{a_p}{a_{p-1}}=(2p-1)\frac{p\left[\frac{\pi\mu}{\sin(\pi\mu)}-1\right]}
{\Gamma(1-\mu)\frac{\Gamma(p\mu+1)}{\Gamma(p\mu+1-\mu)}-1}.
\end{equation}
In the two limiting cases $\mu\rightarrow1^{-}$ (or 
$\beta\rightarrow2^{-}$) and $\mu\rightarrow0^{+}$ (or 
$\beta\rightarrow1^{+}$), Eq.(\ref{momentbis}) gives back 
$a_p/a_{p-1}\rightarrow 2p-1$, namely, the two
Gaussian cases previously mentioned.

\section{Derivation of the results}\label{sec:der}
 
\subsection{Second moment}

\begin{figure}
\begin{center}
\epsfig{figure=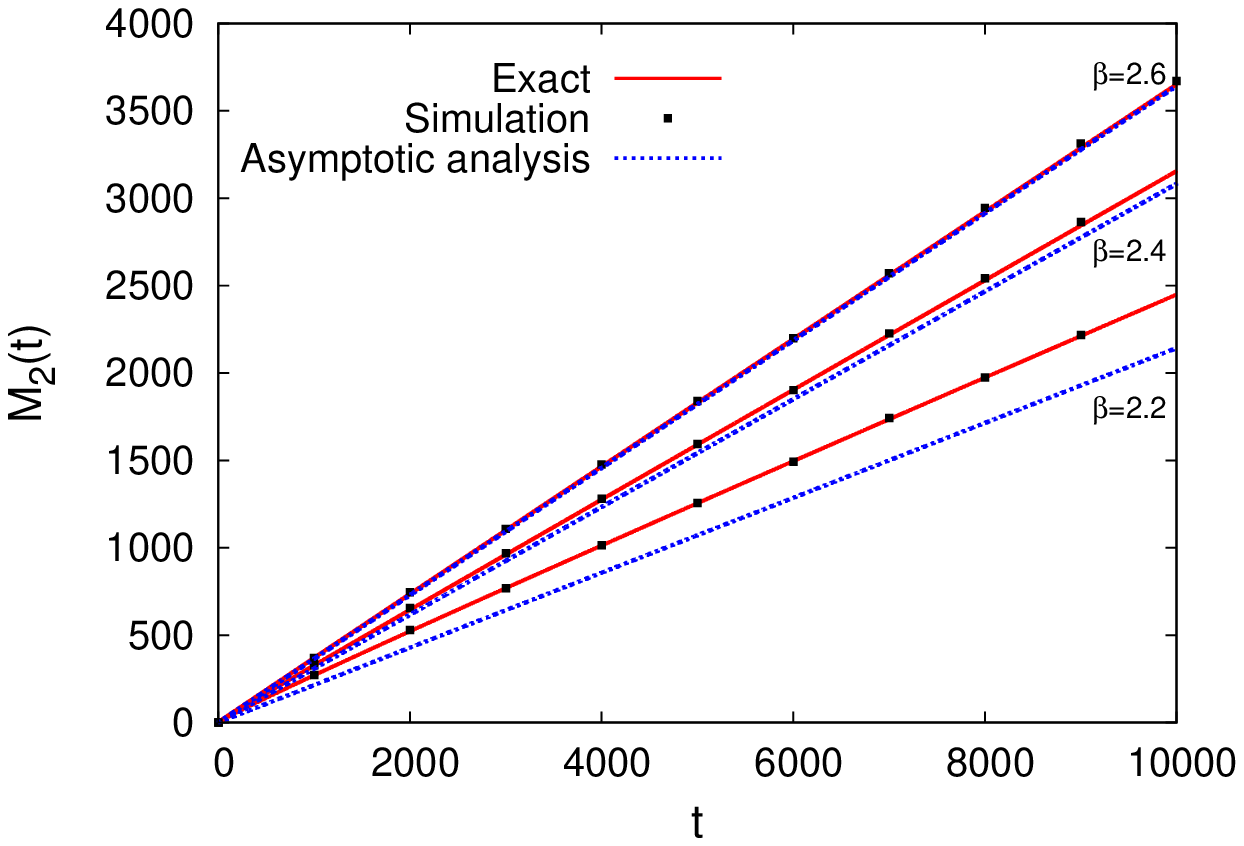,width=3.in}
\epsfig{figure=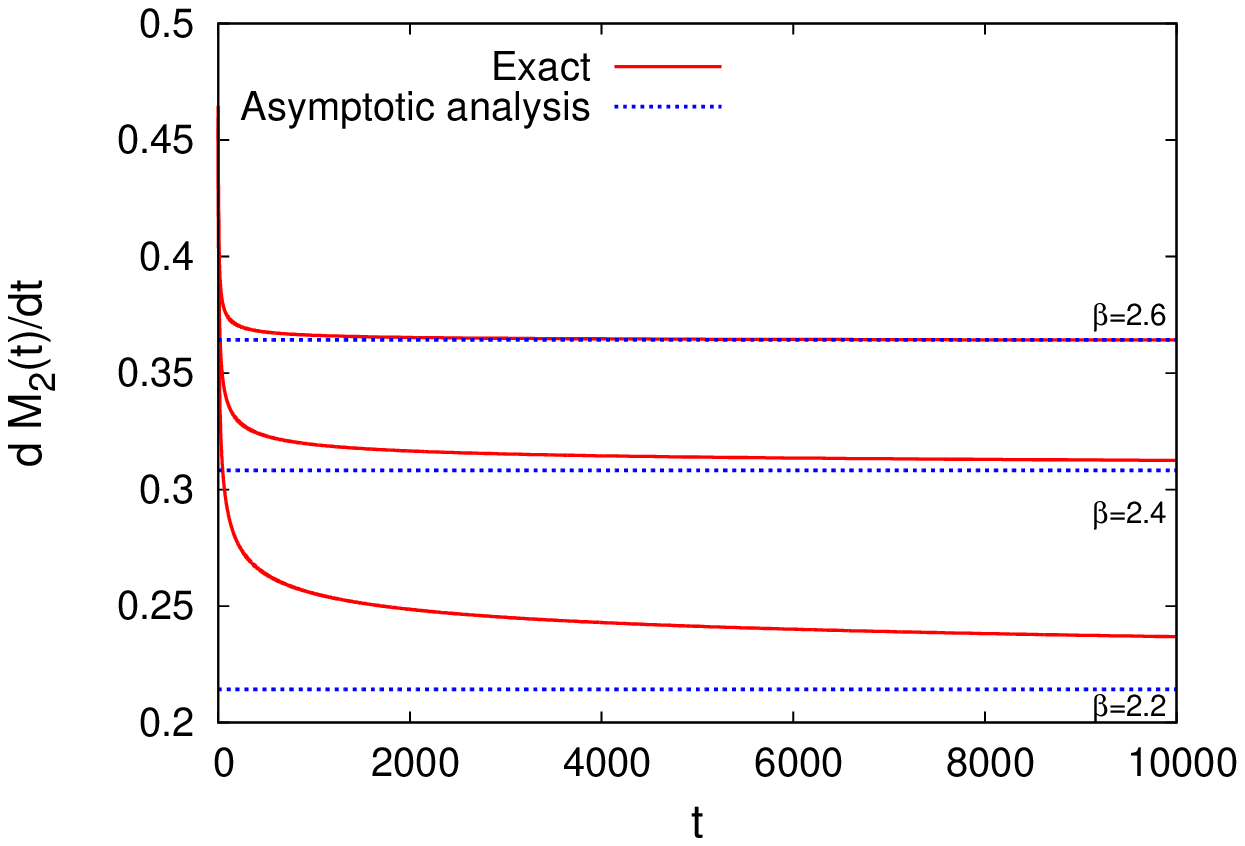,width=3.in}
\end{center}
\vspace{-0.5cm}
\caption{(color online) Comparison between the exact MSD  
obtained from solving Eq. (\ref{genM2}) numerically (solid red lines), and the 
asymptotic analytical results [dotted blue lines, from Eq. (\ref{iv})], 
in the normal diffusive case $\beta>2$. {\bf Top:} $M_2(t)$ vs. $t$. The 
black dots
correspond to Monte Carlo simulations.
{\bf Bottom:} The same curves, represented as $dM_2(t)/dt$ vs. $t$. Higher order 
corrections, which are 
not calculated in the theory, become more important at finite $t$ when $\beta$ 
approaches the subdiffusive transition point $\beta=2$. In all cases, $q=0.5$.}
\label{fig:RelMemlinear}
\end{figure}

By taking the second moment of Eq.(\ref{genreset}), we obtain a recursive 
relation for the mean square displacement:
\begin{equation}\label{genM2}
M_2(t+1)=1-q+(1-q)M_2(t)+\frac{q}{C(t)}\sum_{t'=0}^{t}F(t-t')M_2(t').
\end{equation}
This equation may not be exactly solvable for all $t$, but we
investigate its asymptotic behavior. To show (\ref{iv})-(\ref{vi}), 
we write Eq.(\ref{genM2}) in the form:
\begin{equation}\label{genM2b}
M_2(t+1)-M_2(t)-(1-q)+qM_2(t)=\frac{q}{C(t)}{\cal F}\{M_2(t)\}
\end{equation}
with $C(t)=\sum_{\tau=0}^t(1+\tau)^{-\beta}$ and
\begin{equation}\label{genF}
{\cal F}\{M_2(t)\}\equiv\sum_{t'=0}^{t}\frac{M_2(t')}{(t-t'+1)^{\beta}}.
\end{equation}

We notice that a diverging $M_2(t)$ in the infinite time limit implies
that the terms with large $t'$ dominate in the sum (\ref{genF}). Therefore,
one may approximate $M_2(t')$ in the sum by its leading asymptotic form, even
at small $t'$.

If $\beta>2$, we make the asymptotic ansatz $M_2(t)\simeq K t$, obtaining:
\begin{eqnarray}
{\cal F}\{Kt\}&=&
Kt\sum_{\tau=0}^{t}(1+\tau)^{-\beta}-K\sum_{\tau=0}^{t}\tau(1+\tau)^{-\beta}\\
&\simeq&Kt\sum_{\tau=0}^{\infty}(1+\tau)^{-\beta}-
K\sum_{\tau=0}^{\infty}\tau(1+\tau)^{-\beta},\label{F1}
\end{eqnarray}
since the two sums above converge.
Substituting (\ref{F1}) in (\ref{genM2b}) and noting that
$M_2(t+1)-M_2(t)\simeq K$, one 
obtains an equation for $K$, which is easily solved as
$K=(1-q)/[1+q\langle\tau\rangle_{\infty}]$. This is result (\ref{iv}),
which is displayed in Figure \ref{fig:RelMemlinear} (top).
This panel also shows the MSD obtained 
by iterating the exact Eq. (\ref{genM2}) numerically from the initial 
condition $M_2(0)=0$, as well as the MSD obtained from Monte Carlo 
simulations (both
are identical within numerical errors).
Figure \ref{fig:RelMemlinear} (bottom) actually shows that
the time derivative of the exact MSD tends 
to the calculated diffusivity $K$. Note that higher order corrections to the
linear behavior become more important at finite $t$ as $\beta\rightarrow2$,
and convergence is slower.
At $\beta=2$, $\langle\tau\rangle_{\infty}=\infty$ and the diffusivity $K$ 
vanishes, suggesting a change of temporal behavior.

\begin{figure}
\begin{center}
\epsfig{figure=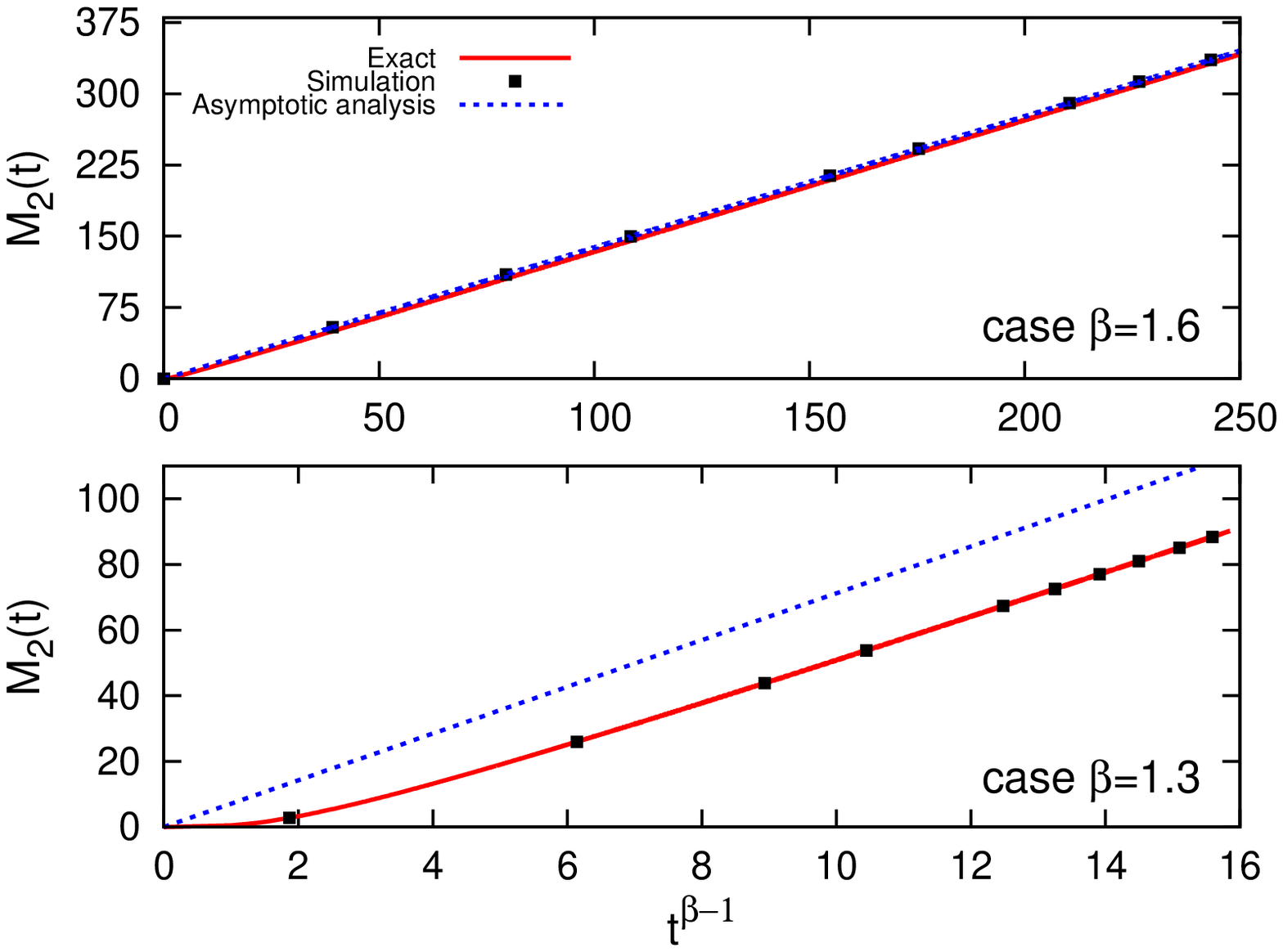,width=3.in}
\epsfig{figure=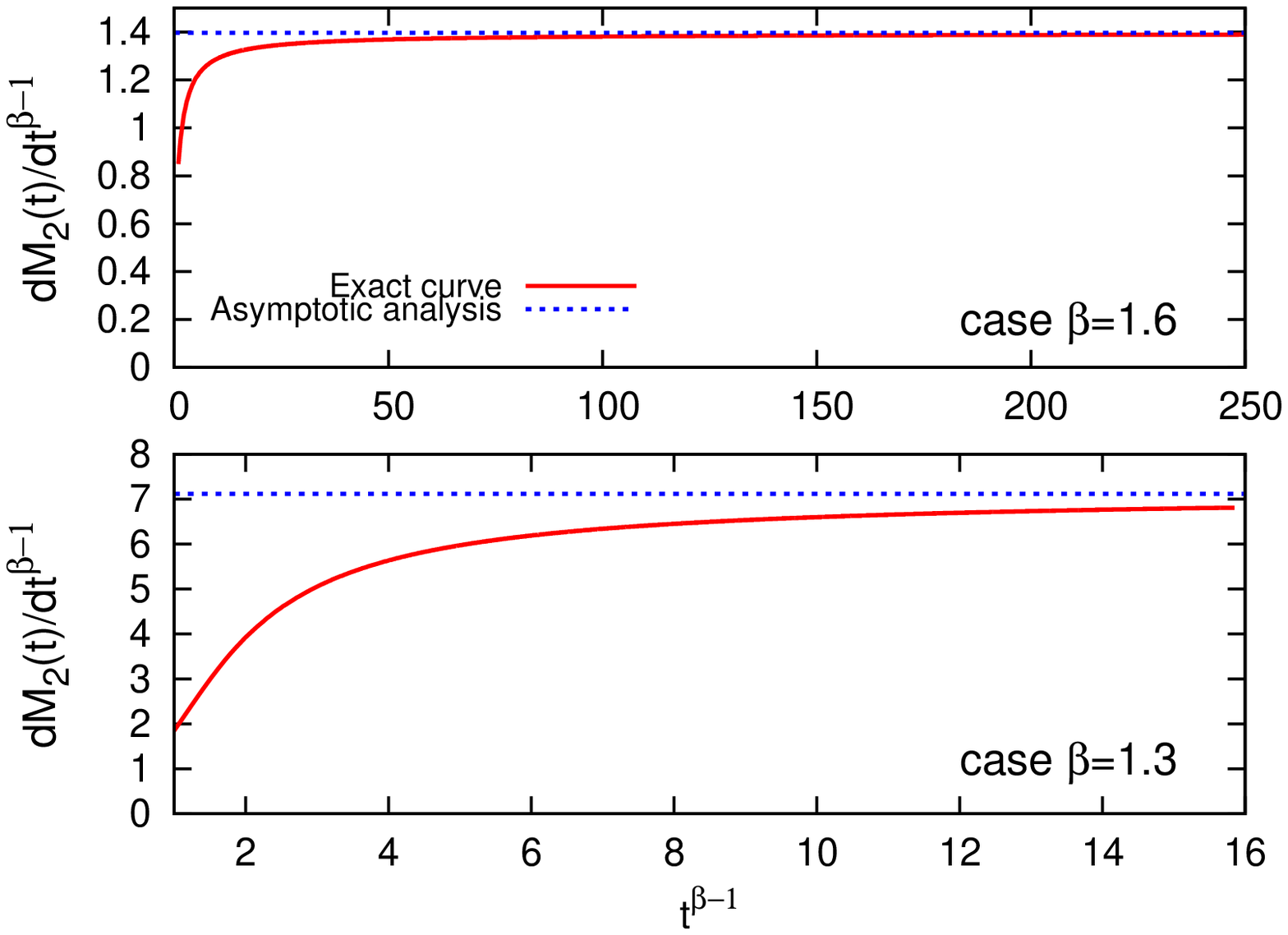,width=3.in}
\end{center}
\vspace{-0.5cm}
\caption{(color online) {\bf Top panels:} $M_2(t)$ as a function of $t^{\beta-1}$ for
two values of $\beta$ in the range $(1,2)$ and with $q=0.5$. The solid red 
lines are obtained from exact resolution of Eq.(\ref{genM2}) and the dotted 
blue lines are given by Eq. (\ref{v}). The black dots correspond to Monte 
Carlo simulations of the model. {\bf Bottom panels:} Same curves, 
represented as $dM_2(t)/d(t^{\beta-1})$ vs. $t^{\beta-1}$.}
\label{fig:RelMemsubdiff}
\end{figure}

To examine the case $1<\beta<2$, we make the general 
ansatz $M_2(t)\simeq K t^{\nu}$,
where $\nu$ and $K$ are unknown. Similarly to (\ref{F1}), we seek to expand 
the memory term as  ${\cal F}\{Kt^{\nu}\}\simeq Kt^{\nu}(c_1+c_2t^{-\beta+1})$
for large $t$. As shown in the Appendix \ref{app_mem}:
\begin{eqnarray}\label{F2}
{\cal F}\{Kt^{\nu}\}&\simeq& Kt^{\nu}
\left\{\zeta(\beta)-t^{-\beta+1}\left[\int_0^{1}du\frac{1-(1-u)^{\nu}}{u^{\beta}}
\right.\right. \nonumber\\
&+&\left.\left.\frac{1}{\beta-1}\right]\right\}.
\end{eqnarray}
The normalization constant $C(t)$ can be written as:
\begin{equation}
C(t)=\sum_{\tau=0}^{t}\frac{1}{(1+\tau)^{\beta}}\simeq
\zeta(\beta)-\frac{t^{-\beta+1}}{\beta-1}.\label{Cta}
\end{equation}
Assuming that $\nu<1$ (subdiffusive behavior), 
so that we can neglect $M_2(t+1)-M_2(t)\simeq\dot{M}_2(t)$ in (\ref{genM2b}) 
compared to the constant $-(1-q)$, and substituting (\ref{F2}) and (\ref{Cta})
in (\ref{genM2b}) we obtain:
\begin{equation}\label{a1}
-(1-q)=qKt^{\nu-\beta+1}\frac{1}{\zeta(\beta)}\int_0^1du
\frac{u^{\nu}-1}{(1-u)^{\beta}}+O(t^{-\beta+1}).
\end{equation}
The exponent of $t$ in the right-hand-side of (\ref{a1}) 
must be 0 for consistency, which gives $\nu=\beta-1\equiv\mu$. 
Consequently, this equation 
can be solved for the constant $K$ and formula (\ref{v}) is obtained.
Figure \ref{fig:RelMemsubdiff} (top) shows that the exact numerical MSD 
obtained from iterating (\ref{genM2}) tends to a linear behavior with respect
to the variable $t^{\beta-1}$, as predicted by our result. Note that
at large but finite time, the exact MSD would be 
better approximated  by the the leading asymptotic result 
plus a constant, which unfortunately cannot be calculated by our method
(this situation occurs for $\beta<1$ as well, see Fig. \ref{fig:RelMemlog} 
below). However, such constant becomes negligible as $t\rightarrow\infty$.
For a more precise comparison, we have displayed in
Figure \ref{fig:RelMemsubdiff} (bottom) the derivative of the MSD
with respect to $t^{\beta-1}$: this quantity tends to a constant in very 
good agreement with the calculated $K$. 

\begin{figure}
\begin{center}
\epsfig{figure=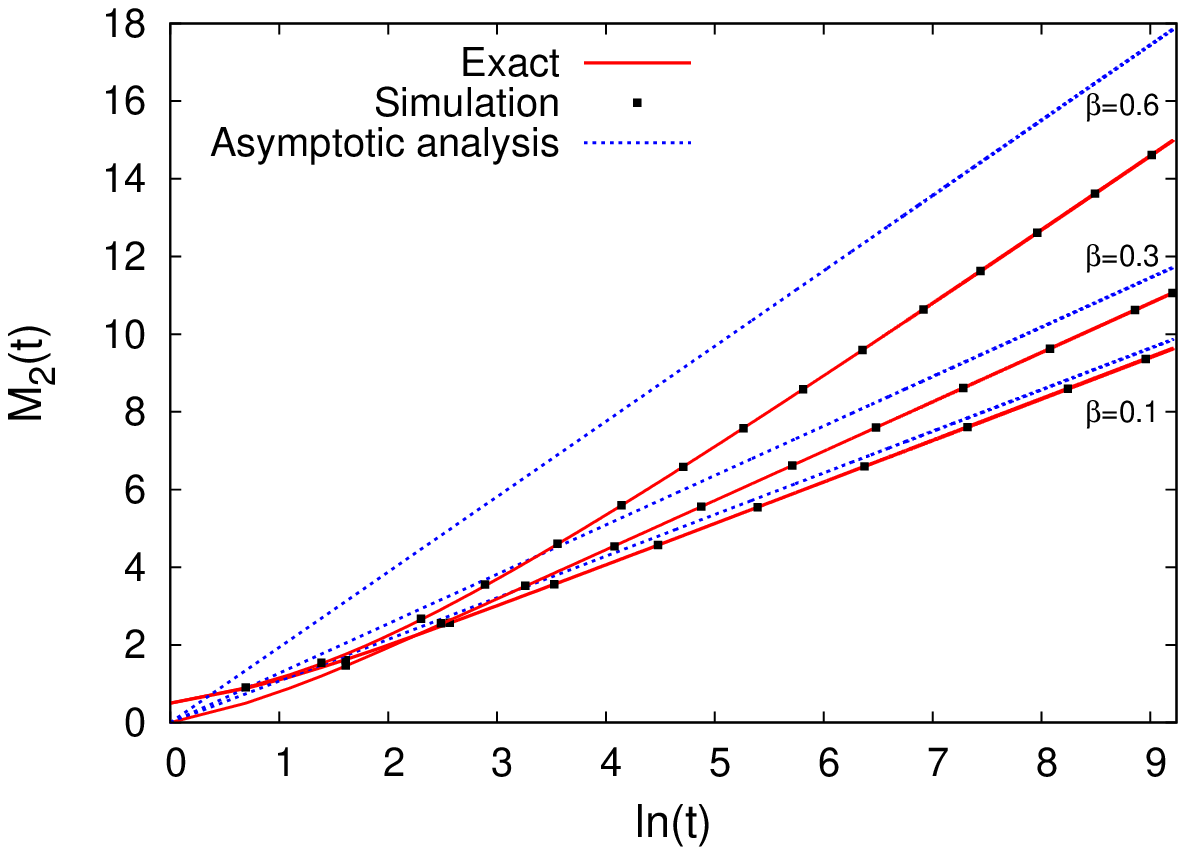,width=3.in}
\epsfig{figure=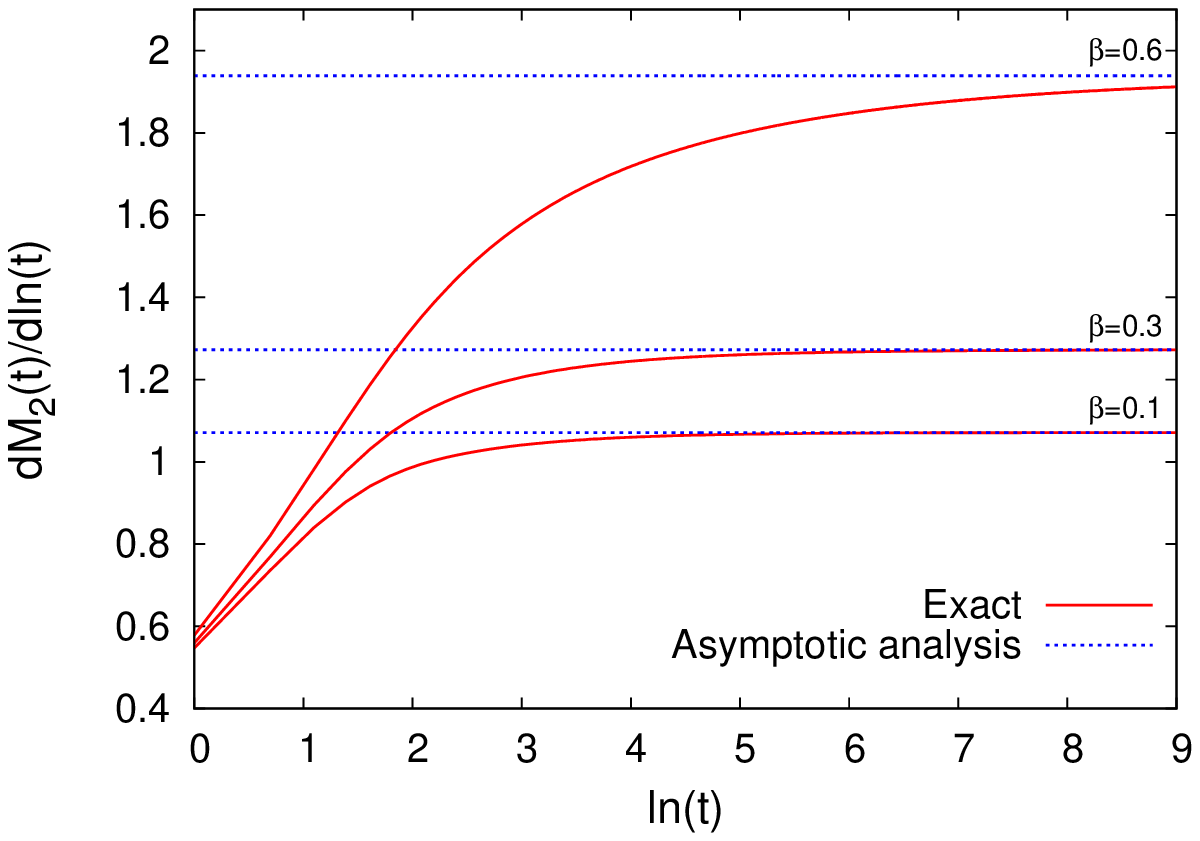,width=3.in}
\end{center}
\vspace{-0.5cm}
\caption{(color online) {\bf Top:} $M_2(t)$ as a function of $\ln t$ for
three values of $\beta<1$ and with $q=0.5$. The solid red lines
are obtained from exact resolution of Eq.(\ref{genM2}) and the dotted blue
lines are given by Eq. (\ref{vi}). The black dots correspond
to Monte Carlo simulations. {\bf Bottom:} The same curves, represented as 
$dM_2(t)/d(\ln t)$ vs. $\ln t$.
}
\label{fig:RelMemlog}
\end{figure}

In the case $\beta<1$, we follow a similar route, now with a logarithmic ansatz
$M_2(t)\simeq K\ln t$, as suggested by the exact result for the 
preferential visit model, {\it i.e.}, $\beta=0$ \cite{boyersolis}. 
Since $\sum_{\tau=0}^t(1+\tau)^{-\beta}$
does not converge, one can use directly the Euler-Maclaurin expansion:
\begin{equation}\label{T4}
C(t)\simeq\int_0^{t}d\tau(1+\tau)^{-\beta}
\simeq\frac{t^{1-\beta}}{1-\beta},
\end{equation}
at large $t$. Similarly,
\begin{equation}
{\cal F}\{M_2(t)\}\simeq\int_0^{t}dt'\frac{M_2(t')}{(t-t'+1)^{\beta}}
\simeq K\int_A^{t}dt' \frac{\ln t'}{(t-t'+1)^{\beta}},
\end{equation}
where $A$ is some constant.
Making the change $t'=ut$ and noticing that the functions
$\ln u/(1-u)^{\beta}$ and $1/(1-u)^{\beta}$ are integrable both 
in $0$ and $1$ when $\beta<1$, we obtain, for $t$ large:
\begin{equation}\label{T5}
{\cal F}\{M_2(t)\}\simeq Kt^{1-\beta}\int_0^1du\frac{\ln u+\ln t}{(1-u)^{\beta}}.
\end{equation}
By substituting (\ref{T4}) and (\ref{T5}) in (\ref{genM2b})
and neglecting $M_2(t+1)-M_2(t)$ as before, we obtain:
\begin{equation}
-(1-q)=qK(1-\beta)\int_0^1 du \frac{\ln u}{(1-u)^{\beta}},
\end{equation}
which gives the constant $K$ and result (\ref{vi}). 
Figure \ref{fig:RelMemlog} (top) shows that the exact MSD obtained
numerically becomes a linear function of $\ln t$ at large times, in agreement 
with theory. We have also displayed in
Figure \ref{fig:RelMemlog} (bottom) the derivative of the MSD
with respect to $\ln t$: this quantity tends to a constant in very 
good agreement with the calculated $K$.

\subsection{Higher order moments}

In a similar way, we now consider the asymptotic behavior of the moments 
$M_{2p}(t)$. We start from the exact relation which follows from the 
master equation (\ref{genreset}):
\begin{eqnarray}\label{m2pexact}
M_{2p}(t+1)&=&1-q+(1-q)M_{2p}(t)\nonumber\\
&+&(1-q)\sum_{k=1}^{p-1}C_{2p}^{2k}M_{2k}(t)\nonumber\\
&+&\frac{q}{C(t)}\sum_{t'=0}^{t}F(t-t')M_{2p}(t').
\end{eqnarray} 
Assuming the scaling hypothesis (\ref{scaling2}), and given that $M_2(t)$ 
always diverges as $t\rightarrow\infty$, we can neglect the 
term $1-q$ as well as the terms proportional to $M_{2k}(t)$ for all 
$k<p-1$ in the right-hand-side of (\ref{m2pexact}). 
Substituting $M_{2p}(t+1)-M_{2p}(t)$ by $dM_{2p}/dt$ in the long time limit, 
Eq. (\ref{m2pexact}) becomes:
\begin{eqnarray}
a_p\frac{dM_2(t)^p}{dt}&\simeq&(1-q)p(2p-1)a_{p-1}M_2(t)^{p-1}\nonumber\\
&+&\frac{qa_p}{C(t)}\sum_{t'=0}^{t}F(t-t')\left[M_2(t')^p-M_2(t)^p\right].
\label{momgenapprox}
\end{eqnarray}
If $1<\beta<2$, we have $M_2(t)\simeq K t^{\beta-1}$ 
and Eq.(\ref{momgenapprox}) reads:
\begin{eqnarray}\label{momrelapprox}
K^pa_p p(\beta-1)t^{(\beta-1)p-1}&\simeq&(1-q)p(2p-1)a_{p-1}K^{p-1}t^{(\beta-1)(p-1)}\nonumber\\
&+&\frac{qa_p K^p}{C(t)}\sum_{t'=0}^t 
\frac{t'^{(\beta-1)p}-t^{(\beta-1)p}}{(t-t'+1)^{\beta}}.
\end{eqnarray}
Using formula (\ref{F2}), we obtain:
\begin{eqnarray}
\sum_{t'=0}^t \frac{t'^{(\beta-1)p}}{(t-t'+1)^{\beta}}&\simeq& 
t^{(\beta-1)p}\times\nonumber\\
& &\left\{\zeta(\beta)-t^{-\beta+1}
\left[\int_0^{1}du\frac{1-u^{(\beta-1)p}}{(1-u)^{\beta}}\right.\right.\nonumber\\
&+&\left.\left.\frac{1}{\beta-1}\right]\right\},\label{sumtp}
\end{eqnarray}
whereas
\begin{eqnarray}
\sum_{t'=0}^t \frac{t^{(\beta-1)p}}{(t-t'+1)^{\beta}}&\simeq&
t^{(\beta-1)p}
\left[\zeta(\beta)-\frac{t^{-\beta+1}}{\beta-1}\right]\label{sumt}.
\end{eqnarray}
Since $1<\beta<2$, then $t^{(\beta-1)p-1}\ll t^{(\beta-1)(p-1)}$ and
the left-hand-side of Eq.(\ref{momrelapprox}) can be neglected. Substituting
(\ref{sumtp}) and (\ref{sumt}) in (\ref{momrelapprox}), then taking the
limit $C(t)\rightarrow\zeta(\beta)$ at large $t$ and using the expression 
(\ref{v}) for $K$, Eq. (\ref{nongaussian}) is obtained.

For the case $\beta<1$, or $M_2(t)\simeq K\ln t$, the derivative in the
left-hand-side of (\ref{momgenapprox}) is $O(\ln^{p-1}(t)/t)\rightarrow 0$ 
and can still be neglected compared with terms growing with time in 
this equation. Thus Eq.(\ref{momgenapprox}) becomes:
\begin{equation}\label{momrellogapprox}
(1-q)p(2p-1)a_{p-1}K^{p-1}\ln^{p-1}t=
\frac{qa_pK^p}{C(t)}\int_0^1dt' \frac{\ln^p t-\ln^p t'}{(t-t'+1)^{\beta}},
\end{equation} 
where in this case the diverging sums have been replaced by 
integrals. We next write:
\begin{eqnarray}
\int_0^t dt'\frac{\ln^p t'}{(t-t'+1)^{\beta}}&\simeq&
t^{1-\beta}\int_0^1du \frac{(\ln u+\ln t)^p}{(1-u)^{\beta}}\nonumber\\
&=&t^{1-\beta}\sum_{k=0}^pC_p^k\ln^{p-k}t\int_0^1du
\frac{\ln^k u}{(1-u)^{\beta}}\nonumber\\
&\simeq& t^{1-\beta}\left[\frac{\ln^p t}{1-\beta}\right.\nonumber\\
&+&\left. p\ln^{p-1}t
\int_0^1du\frac{\ln u}{(1-u)^{\beta}}\right],\label{sumlogtp}
\end{eqnarray}
where in the last step we have retained only the first two dominant terms 
of the binomial expansion, {\it i.e.} those of indices $k=0$ and $k=1$.
We also have: 
\begin{equation}
\int_0^t dt'\frac{\ln^p t}{(t-t'+1)^{\beta}}\simeq
t^{1-\beta}\frac{\ln^p t}{1-\beta}.\label{sumlogt}
\end{equation}
Substituting  (\ref{sumlogtp}) and (\ref{sumlogt}) 
in (\ref{momrellogapprox}), then using the expansion
$C(t)\simeq t^{1-\beta}/(1-\beta)$ as well as the expression
(\ref{vi}) for $K$, one obtains the simple result: 
\begin{equation}
a_{p}/a_{p-1}=2p-1,
\end{equation} 
which corresponds to the Gaussian distribution.

Note that the equation that leads to this Gaussian result
[Eq. (\ref{momrellogapprox})] can be understood as a
balance between diffusion due to random increments (the combinatorial
left-hand-side) 
and confining effects due to recurrent memory (the integral right-hand-side).
This balance arises because the term $a_pdM_{2}(t)^p/dt$ in Eq. 
(\ref{momgenapprox}) is negligible. 
In the simple random walk, on the contrary, $a_pdM_{2}(t)^p/dt$ is 
dominant, since $M_2(t)\propto t$, and equals the combinatorial
term (in the absence of a memory term). Therefore,
the mechanism leading to Gaussianity in the memory walk here is 
very different from the Central Limit Theorem case.

\subsection{Numerical tests of the higher order moments}

\begin{figure}
\begin{center}
\epsfig{figure=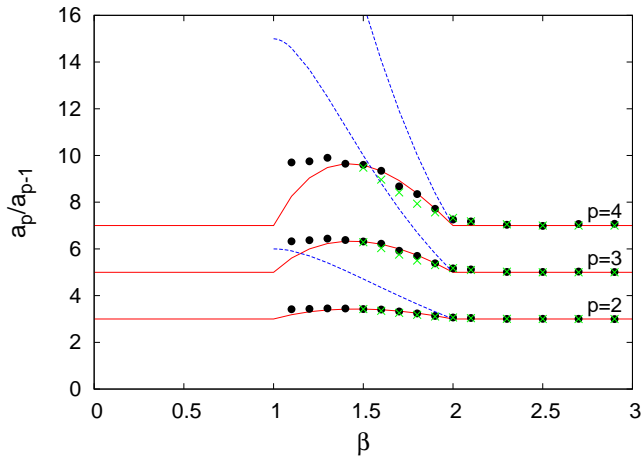,width=3.5in}
\end{center}
\vspace{-0.5cm}
\caption{(color online) Moment ratios $a_p/a_{p-1}$ (for $p=2$, 3 and 4)
as a function of $\beta$. The solid red lines correspond to the 
analytical results.
The two Gaussian regimes ($a_p/a_{p-1}=cst=2p-1$) are
separated by non-Gaussian scalings in the interval $1<\beta<2$, given
by Eq. (\ref{nongaussian}). The dots correspond to the quantity  $Q_p(t)$ 
computed from simulations ($q=0.1$, $t=10^6$, $5\ 10^4$ runs).
The green crosses correspond to $Q_p(t)$ 
computed from exact numerical resolution of Eq.(\ref{genreset}),
with $q=0.1$ and $t=10^4$.
The ratios $a_p/a_{p-1}$ corresponding to the CTRW model (dotted blue lines) 
are shown for comparison for the same values of $p$. In that case, $\beta$ is
the waiting time exponent.}
\label{fig:RelMemmoments}
\end{figure}

Figure \ref{fig:RelMemmoments} shows several analytical curves $a_p/a_{p-1}$
as a function of $\beta$, summarizing the 3 regimes discussed above. 
To check these results, we performed
Monte Carlo simulations of the model and computed the quantity
\begin{equation}\label{Qp}
Q_p(t)\equiv \frac{M_{2p}(t)}{M_{2p-2}(t)M_2(t)}, 
\end{equation}
after obtaining the corresponding moments by averaging $X_t^{2p}$, 
$X_t^{2p-2}$ and $X_t^2$, over many independent walks.
If the scaling hypothesis holds, $Q_p(t)\rightarrow a_p/a_{p-1}$.
In each case, we checked that $Q_p(t)$ tended to a constant at large times, 
which is indicated by a dot in Figure \ref{fig:RelMemmoments}.

The agreement between theory and simulations is very good.
Note, however, that when $\beta$ 
becomes small ($\beta<1.3$), the computed $Q_p(t)$ tends to be larger
than the theoretical value. This is because diffusion is very slow in 
these cases and therefore the scaling regime becomes very
difficult to reach in finite time simulations. Recall that the scaling regime 
settles only when $M_2(t)\gg1$.
In ref. \cite{boyersolis}, the approach toward
the scaling regime was studied in details for the case $\beta=0$. It was shown
that the first correction to scaling slowly decayed
as $1/\ln t$ and thus could be neglected only at extremely long times,
which cannot be reached in practice in simulations.

As an additional test, we solved Eq.(\ref{genreset}) numerically,
calculated the moments from $P(n,t)$, and then calculated $Q_p(t)$.
The results are indicated by crosses in Fig. \ref{fig:RelMemmoments}.
The necessary computer memory scales as $t^2$ in this method,
and times beyond $t=15,000$ could not be reached in practice 
(this is a low value, compared to $t=10^6$ used in 
simulations). The scaling regime is thus less easy to observe, specially in 
the subdiffusive case, and the discrepancy with the 
theoretical $a_p/a_{p-1}$ is a bit larger than in simulations.

\section{Comparison of the case $1<\beta<2$ with the CTRW}

\subsection{Qualitative analogy}

The subdiffusive law (\ref{v}) is the consequence of 
a diverging average backward time $\langle\tau\rangle_t$ as 
$t\rightarrow\infty$. Interestingly, this situation is analogous to that of 
the continuous time random walk (CTRW), where subdiffusion arises 
due to diverging waiting times \cite{shlesinger,kotulski}. In this well-known 
renewal Markovian process, an unbiased random walker remains at its current 
location during a random time $\tau$, i.i.d. from a distribution $\psi(\tau)$, 
before performing the next step to a nearest-neighbor site \cite{montroll}. 
If $\psi(\tau)\sim \tau^{-\beta}$ with 
$1<\beta<2$ (or, setting $\beta=1+\mu$, if $0<\mu<1$), then
$\langle\tau\rangle=\infty$ and a diffusion constant cannot be properly 
defined. In this case, motion is subdiffusive with $M_2(t)\sim t^{\beta-1}$, 
the same scaling as in Eq. (\ref{v}).

\begin{figure}
\begin{center}
\epsfig{figure=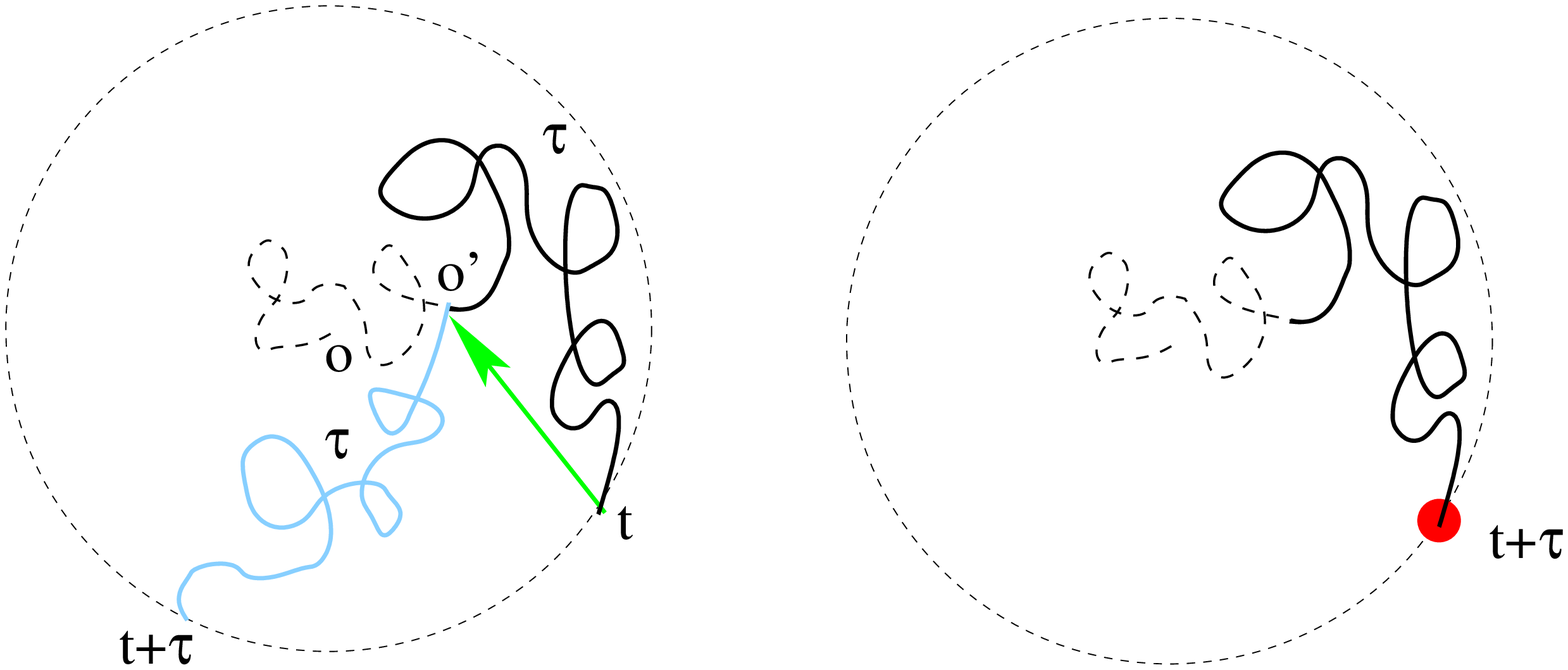,width=3.2in}
\end{center}
\vspace{-0.5cm}
\caption{Qualitative comparison between the current model (left) 
and a CTRW (right), see text.}
\label{fig:RelMemCTRW}
\end{figure}

The following qualitative argument can be made to explain the similar 
behaviors of the mean square displacements in these two models. Let us consider a
walker in our model starting from the origin $O$ and taking a relocation 
step at a time $t$ (Figure \ref{fig:RelMemCTRW}, left). This step, represented 
by the arrow, brings the walker back to some position $O'$ that was occupied 
earlier, $\tau$ time units before time $t$, where $\tau$ is a random
variable $\le t$ drawn from the p.d.f. $F(\tau)\propto \tau^{-\beta}$. 
Just before taking this step, 
the walker
was at some distance from the origin, represented by the large circle. In order
to reach this circle again from $O'$, the walker will typically need to
diffuse during another time interval $\tau$ again (assuming stationarity in 
the process). Hence, only at time about $t+\tau$ the walker's
displacement may grow beyond the radius of the circle. This situation can be 
seen as analogous to keeping the walker immobile at the position 
reached at $t$ during a time $\tau$, which is the basic rule of the CTRW 
(Figure \ref{fig:RelMemCTRW}, right).

However, one may intuitively guess that the two models differ quantitatively, 
namely, that their probability densities $P(n,t)$ are given 
by different scaling functions $g(x)$. This is actually the case, since 
$g(x)$ is exponential when $\beta\rightarrow1$ in the one-dimensional
CTRW, {\it i.e.}, 
$g^{CTRW}(x)\rightarrow \frac{1}{\sqrt{2}}\exp(-\sqrt{2}|x|)$ \cite{bouchaud}, 
instead
of the Gaussian form in the memory model. More generally, 
for $1<\beta<2$ the expression of $g^{CTRW}(x)$ is exactly known in terms of 
the fully asymmetric L\'evy law of index $(\beta-1)/2$. Its moments can 
be calculated exactly, giving \cite{bouchaud}:
\begin{equation}\label{momentsCTRW}
\left.\frac{a_p}{a_{p-1}}\right|_{CTRW}=(2p-1)(p-1)
\frac{\Gamma[\mu(p-1)]\Gamma(\mu+1)}{\Gamma(\mu p)}.
\end{equation}
The above expression is indeed different from (\ref{nongaussian})
or (\ref{momentbis}).
The limit distribution $g(x)$ in the present model does not reduce, 
to our knowledge, to a standard distribution. Figure \ref{fig:RelMemmoments} 
displays $a_p/a_{p-1}$ as a function of $\beta$ for $p=2,3$ and $4$ for the
CTRW and the present model, in which deviations from Gaussianity are weaker.

\subsection{Large $x$ behavior of $g(x)$}

Further insights on $g(x)$ in the memory model can be gained by studying its 
behavior at large $x$ for $\beta$ in the range $(1,2)$. This asymptotic 
behavior can be derived by analyzing the moments in the limit $p\gg 1$.
Knowing the properties of $g(x)$ is also useful to understand how the 
distribution tends to a Gaussian instead of a exponential as 
$\beta\rightarrow1^+$. 

The large $p$ behavior of (\ref{momentbis}) for any $0<\mu<1$ 
can be obtained by using the Stirling formula 
$\Gamma(az+b)\simeq\sqrt{2\pi}e^{-az} (az)^{az+b-1/2}$ for large $z$:
\begin{equation}\label{nongaussianlargep}
\frac{a_p}{a_{p-1}}\simeq\chi_{\mu}\ p^{2-\mu}\quad{\rm with}\quad 
\chi_{\mu}=2\frac{\frac{\pi\mu}{\sin(\pi\mu)}-1}{\Gamma(1-\mu)\mu^{\mu}}.
\end{equation}
Clearly, the asymptotic result (\ref{nongaussianlargep}) is valid as long
as $\chi_{\mu}$ is non-zero. This prefactor is strictly positive if 
$0<\mu\le 1$ but vanishes at $\mu=0$ and this case will deserve a 
special attention below.
 
If $0<\mu<1$, equation (\ref{nongaussianlargep}) implies that the 
scaling function follows at large $x$ the leading behavior: 
\begin{equation}\label{g}
g(x)\propto e^{-b_{\mu} |x|^{\delta_{\mu}}},\quad |x|\gg1, 
\end{equation}
with 
\begin{equation}\label{cstg}
\delta_{\mu}=\frac{1}{1-\mu/2}\quad{\rm and}\quad 
b_{\mu}=2\chi_{\mu}^{-\frac{\delta}{2}}/\delta_{\mu}.
\end{equation}
Let us notice that the large $x$ behavior above 
is similar to that of the CTRW with a waiting time distribution 
$\psi(\tau)\sim \tau^{-(1+\mu)}$. 
Namely, the leading part of $g^{CTRW}(x)$ is also of the form
(\ref{g}) and with the same exponent $\delta_{\mu}$ as given by (\ref{cstg}). 
This property stems 
from the fact that the moment relation for CTRW, Eq.(\ref{momentsCTRW}), 
also becomes proportional to $p^{2-\mu}$ at large $p$, like in 
(\ref{nongaussianlargep}). However, the scaling functions do differ in 
the two models since the prefactors $\chi_{\mu}$ (and therefore $b_{\mu}$ 
in the exponential) are different.

This difference increases drastically as $\mu\rightarrow0$: 
it is easy to see from (\ref{nongaussianlargep}) that 
\begin{equation}
\chi_{\mu}\rightarrow0\quad {\rm as}\quad \mu\rightarrow0
\end{equation} 
whereas $\chi_{\mu=0}^{CTRW}$ is $>0$. The fact that 
the prefactor vanishes is a crucial property of the memory model. It indicates 
that the leading term of the scaling function at large $x$ is no longer given 
by (\ref{g})-(\ref{cstg}). A higher order calculation 
shows that, for any $p$:
\begin{equation}
\frac{a_p}{a_{p-1}}\rightarrow 2p-1 \quad {\rm as}\quad \mu\rightarrow0^+,
\end{equation}
which confirms our previous result that $g(x)$ becomes Gaussian again at $\mu=0$. An expansion of 
the moment relation (\ref{momentbis}) at large $p$ and small $\mu$ 
generalizes Eq.(\ref{nongaussianlargep}):
\begin{equation}
\frac{a_p}{a_{p-1}}\simeq \frac{\pi^2\mu^2}{3} p^{2-\mu}+2p-1,\quad
(p\gg1,\ \mu\ll1).
\end{equation}
Therefore the exponential behavior of $g(x)$  at large $x$ 
({\it i.e.} $a_p/a_{p-1}\propto p^2$ for large $p$) is avoided at $\mu=0$ 
due to the vanishing prefactor and the distribution adopts a Gaussian profile 
instead.

\section{L\'evy-like distributions of relocation lengths}\label{sec:Levy}

Every time the walker uses memory, it performs a relocation step, of length, 
say, $\ell$, corresponding to the distance between its current location 
and the newly chosen site. 
These lengths $\ell$ are represented by the thick (black)
lines in the one dimensional simulations of Figure \ref{fig:RelMemTraj}. They
can be quite large in the subdiffusive regime. We present below 
a scaling argument to obtain an expression for the probability 
distribution of $\ell$. 

\begin{figure}
\begin{center}
\epsfig{figure=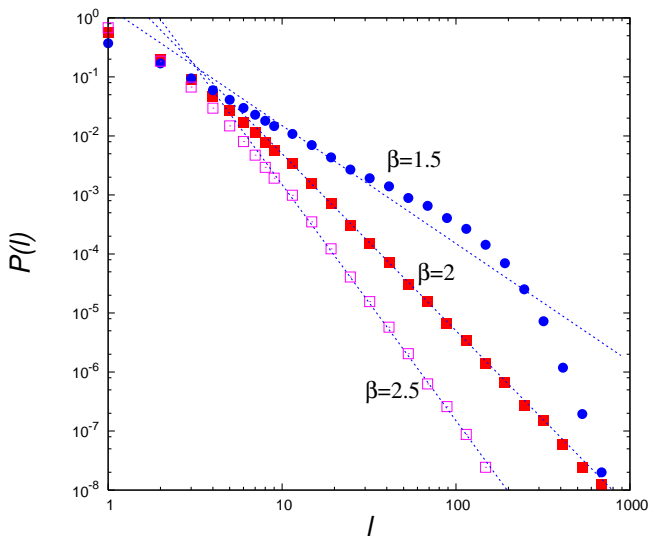,width=2.8in,angle=-90}
\end{center}
\vspace{-0.5cm}
\caption{Probability distribution function of the relocation length
$\ell$ in simulations for several values of $\beta$ ($q=0.01$ in all cases). The solid 
lines have slopes 
$-(2\beta-1)$, {\it i.e.}, $-4$, $-3$ and $-2$ from bottom to top. The
variance of the distribution thus becomes infinite at the subdiffusion 
threshold $\beta=2$.}
\label{fig:pl}
\end{figure}

Since the site occupied after a relocation is the site that was 
occupied $\tau$ time units ago, one may expect that, typically,
\begin{equation}\label{elltau}
\ell\sim [M_2(\tau)]^{1/2}.
\end{equation}
In the case with $\beta>2$, we know that $M_2(t)\propto t$ 
thus the random variables $\ell$ and $\tau$ should be related through the
scaling relation $\ell\sim \tau^{1/2}$. Denoting $P(\ell)$ the
distribution function of $\ell$, probability conservation 
implies $P(\ell)d\ell=\psi(\tau)d\tau$, where $\psi(\tau)\sim \tau^{-\beta}$. 
Hence
\begin{equation}\label{pell}
P(\ell)\sim \ell
\left(\ell^{2}\right)^{-\beta}\sim \ell^{-(2\beta-1)}.
\end{equation}
Thus, when $\beta=2$, the distribution 
$P(\ell)$ obeys a \lq\lq L\'evy-like" 
power-law distribution with marginal exponent $2\beta-1=3$, which
is characterized by an infinite variance $\langle\ell^2\rangle$. 
If $\beta>2$, Eq. (\ref{pell}) indicates that the step length distribution 
exponent is $>3$, outside the L\'evy range, and $\langle \ell^2\rangle$
is therefore finite. 
Note that, unlike in a genuine L\'evy process, the steps in the present 
model are long-range correlated and thus not independent.
Consequently, our model exhibits a paradoxical
behavior: the divergence of the variance of the relocation length $\ell$ 
happens at the onset 
of {\it sub}diffusion, and not superdiffusion as in usual L\'evy flights. 
This important feature can be understood by the fact that 
when the relocation lengths become large, they also become strongly 
anticorrelated and thus limit the walker's diffusion overall. 
The scaling law predicted in (\ref{pell}) agrees very well with 
simulation results, as shown in Figure \ref{fig:pl}.

The case $\beta\in(1,2)$, corresponding to subdiffusive motion, is more 
complicated and the above argument to obtain  $P(\ell)$ is not applicable
strictly speaking. 
Eq.(\ref{elltau}) may not hold due to non-ergodic effects, which are
well known to be strong in subdiffusive processes such as the CTRW
\cite{he,barkai}. 
If we naively assume that (\ref{elltau}) is correct (which makes the implicit 
assumption that the mean square displacement of the walker during an interval 
of time $\tau$ is independent of the time taken as the origin along 
the trajectory), 
we have $\ell\sim \tau^{({\beta-1})/2}$ from  (\ref{elltau}), which gives
$P(\ell)\sim \ell^{\frac{2}{\beta-1}-1}[\ell^{\frac{2}{\beta-1}}]^{-\beta}
\sim\ell^{-3}$. Hence, the relocation length distribution would
remain at the border of the L\'evy range for any $1<\beta<2$.

But this prediction is not confirmed by the numerical results
of Figure \ref{fig:pl}, which show that for intermediate values of $\ell$ 
the distribution $P(\ell)$ decays more slowly than the law $\ell^{-3}$. 
If we assume 
that the average displacement $\ell$ 
taken along the trajectory looks normally diffusive, like in the ordinary CTRW 
process \cite{sokolov,he}, then we should use the scaling $\ell\sim \tau^{1/2}$ 
instead of $\tau^{({\beta-1})/2}$ .
This relation leads to the same form (\ref{pell}) obtained for 
$\beta>2$. Thus the step length exponent would be $2\beta-1$, spanning 
the whole L\'evy range $(1,3)$ as $\beta$ is varied is between 1 and 2. 
The numerical simulations support this qualitative prediction partly. 
As displayed by Figure \ref{fig:pl}, $P(\ell)$ exhibits an intermediate 
power-law regime with an exponent quite close to $2\beta-1$ (case $\beta=1.5$). 
Nevertheless the actual distribution seems more complicated
than an inverse power-law and it is truncated  
at large $\ell$.

\section{Case $\beta<1$ and the Scaled Brownian Motion}

Scaled Brownian motion (SBM) is a well-known anomalous diffusion
process with Gaussian property. Typically, it is simply generated by rescaling 
the time variable of an ordinary Brownian Motion as $t\rightarrow t^{\alpha}$,
with $0<\alpha<2$ a constant (see, {\it e.g.}, \cite{effecFP}). 
Thus, the diffusion front $P(n,t)$ of a SBM is Gaussian, but with 
variance $2D_0t^{\alpha}$, instead of $2D_0t$ in the original process. 
Consequently, it obeys a Fokker-Planck equation (in continuous space)
of the form:
\begin{equation}\label{SBM}
\frac{\partial P}{\partial t}=D(t)\frac{\partial^2 P}{\partial n^2}
\end{equation}
with $D(t)$ a time-dependent diffusion coefficient, given by 
$D(t)=\alpha D_0 t^{\alpha-1}$. 

When $\beta<1$ in the present model, and in the very long time limit,
$P(n,t)$ becomes a Gaussian, too, but with variance 
$M_2(t)\simeq K_{\beta}\ln t$, where the generalized mobility $K_{\beta}$
is given by Eq. (\ref{vi}). 
Therefore, setting $T=\ln t$ we see that $P(n,T)$ obeys the simple 
effective diffusion equation: 
$\partial_T P\simeq \frac{K_{\beta}}{2}\partial_{n^2}P$, in the
limit $T\gg 1$. Coming back to the time variable, the
equation becomes:
\begin{equation}
\frac{\partial P}{\partial t}\simeq\frac{K_{\beta}}{2t}
\frac{\partial^2P}{\partial n^2},
\end{equation}
which is a SBM in the limit $\alpha\rightarrow0$,
{\it i.e.}, with $D(t)\propto 1/t$ at large times. This result
highlights the non-stationary nature of the process due 
to the very long-ranged memory, which gradually slows down the walker.
This mapping is also interesting because it provides a concrete
example of a random walk with constant parameters and constant time increment 
that can be described as a SBM asymptotically (at least, as far 
as the probability function $P(n,t)$ is concerned). Here, 
the SBM time rescaling is not of the power-law form usually considered
but is given by
\begin{equation}
t\rightarrow \ln t.
\end{equation}
Conversely, this type of SBM can be 
considered as a mean-field model for the path-dependent random walks
studied here (individual trajectories clearly differ in the two 
processes).
As noted earlier, we emphasize that this correspondence 
is only valid at very large times. At moderate $\ln t$ ({\it e.g.}, for
$t\sim 10^{10}$ or any numerical simulation time in practice), the 
memory model does not obeys simple scaling {\it a priori}.

The SBM analogy could be used to infer the first-passage 
properties of the model (although they would be limited
to the regime of extremely long times). In addition, the non-ergodic 
properties of SBM systems have been recently studied in details 
in \cite{theil2014,jeong2014}, and could find an application here.

\section{Conclusions}

Non-Markovian, path-dependent processes are in general difficult
to analyze for a single particle, since a rigorous description 
usually requires the
introduction of multiple-time distribution functions, that are
related to each other via a hierarchy of master equations \cite{peliti}. 

Here we have introduced a path-dependent random walk model
that does not exhibit such complications and can be described
by the single-time distribution function. This property
stems from the linear nature of the model, in which the walker
can perform non-local steps backwards in time. 
With this form of self-attraction, the process can be exactly described by one 
master equation. Additionally, the equation is solved exactly in the long 
time limit thanks to the scaling hypothesis. Therefore, a rather precise
picture of the dynamics is obtained, which is not limited to the derivation
of the MSD.

The model exhibit three regimes, depending on how fast memory decays.
In the weakly non-Markovian regime, diffusion is asymptotically normal
but with a reduced diffusion coefficient. A transition to a second regime 
occurs if the mean memory time diverges, {\it i. e.} when the memory 
kernel decays as $\tau^{-\beta}$
with $\beta=2$. In this case, recurrence to previously visited sites 
increases sharply and motion becomes subdiffusive. 
When memory decays very slowly, slower than $\tau^{-1}$, a third, \lq\lq
ultra-slow" diffusion regime with logarithmic behavior settles. 
 
The diffusion front is a function which is practically unknown for reinforced 
walk models, and almost never studied for other non-Markovian models which are
solvable for the MSD \cite{elephant1,elephant2,italian} (but see
\cite{gandhiPRE} for a discussion on the so-called Elephant Walk).
The knowledge of the scaling function here has allowed us to establish 
useful connections between a reinforced walk and 
well-known Markovian models of anomalous diffusion: namely,
the CTRW ($1<\beta<2$) and the SBM ($\beta<1$). The present 
model exhibits distinctive features, such as a new
scaling function defined 
through Eq. (\ref{momentbis}), and anti-correlated L\'evy steps with 
exponent $-3$ at the subdiffusive transition.
The visitation and first-passage statistics are also likely to be
peculiar in this model, and their study certainly deserves future work.

Our results also unveil
a novel, memory driven mechanism for the emergence of L\'evy flights. 
Whereas L\'evy flights are a paradigmatic model of superdiffusive
behavior in physical \cite{bouchaud} and biological \cite{randomsearch} 
systems, the model exposed here illustrates that (truncated) L\'evy flights 
can be also closely linked to subdiffusion. This aspect is 
particularly relevant to the mobility of many living organisms like
humans \cite{geisel,gonzalez,song,rhee} or non-human primates \cite{interface}, 
to name a few, 
who often combine two apparently contradictory patterns: a
power-law step length distribution and a very slow diffusion or 
limited space use.

\begin{acknowledgments}
We thank L. Lacasa, H. Larralde, F. Leyvraz, I. P\'erez-Castillo, 
A. Robledo, S. Thurner for fruitful discussions and D. Aguilar for 
technical support. JCRRC acknowledges supports from CONACYT, PAEP (Posgrado
en Ciencias F\'\i sicas, UNAM) and the Universidad Aut\'onoma de la
Ciudad de M\'exico.
This work was also supported by the PAPIIT Grant IN101712 of the Universidad 
Nacional Aut\'onoma de M\'exico.
\end{acknowledgments}

\appendix

\section{Derivation of the master equation (\ref{genreset})}\label{master}

Let $X_t$ be the position of the walker at time $t$ and let define
$P(n,t+1|i_{t},i_{t-1},...,i_0)$ the probability that $X_{t+1}=n$,
given that $\{X_0,X_1,...,X_t\}=\{i_0,i_1,...,i_t\}$, a prescribed set 
of integers. The model is defined by:
\begin{eqnarray}\label{eq1}
P(n,t+1|i_{t},i_{t-1},..,i_0)&=&\frac{1-q}{2}\delta_{i_t,n-1}
+\frac{1-q}{2}\delta_{i_t,n+1}\nonumber\\
&+&q\sum_{t'=0}^t p_t(t')\delta_{i_{t'},n},
\end{eqnarray}
where the last non-Markovian term indicates that $n$ can be revisited 
if it has been visited at any previous time $t'$. In this memory movement 
mode, which is used with probability $q$, site
$n$ does not need to be a nearest-neighbor of the current position $i_t$, 
and the probability that $n$ is chosen depends on all its previous visits
weighted with the memory kernel $p_t(t')$.

Let define $p(i_1,...,i_t|i_0)$ the probability of a particular trajectory 
generated with the model rules and starting at $i_0$. 
We have the general relation:
\begin{equation}\label{eq2}
\sum_{i_1}...\sum_{i_t}P(n,t+1|i_{t},..,i_0)p(i_1,..,i_t|i_0)
=P(n,t+1|i_0),
\end{equation}
where $P(n,t+1|i_0)$ is the single-time occupation probability, evaluated at 
time $t+1$. In addition, 
for any integer $t'$ in $[0,t]$:
\begin{equation}\label{eq3}
\sum_{i_1}...\sum _{\substack{
i_k\\ k\neq t'}}...\sum_{i_t} p(i_1,...,i_t|i_0)=
P(i_{t'},t'|i_0).
\end{equation}
Multiplying Eq. (\ref{eq1}) by  $p(i_1,...,i_t|i_0)$ and 
summing over all possible values of $i_1$,..,$i_t$, 
we obtain, using Eqs. (\ref{eq2}) and (\ref{eq3}):
\begin{eqnarray}
P(n,t+1|i_0)&=&\frac{1-q}{2}\sum_{i_t}\delta_{i_t,n-1}
P(i_t,t|i_0)\nonumber\\
&+&\frac{1-q}{2}\sum_{i_t}\delta_{i_t,n+1}
P(i_t,t|i_0)\\
&+&q\sum_{t'=0}^{t}p_t(t')\sum_{i_{t'}}\delta_{i_{t'},n}
P(i_{t'},t'|i_0),\nonumber
\end{eqnarray}
or:
\begin{eqnarray}
P(n,t+1|i_0)&=&\frac{1-q}{2}P(n-1,t|i_0)+\frac{1-q}{2}P(n+1,t|i_0)\nonumber\\
&+&q\sum_{t'=0}^{t}p_t(t')P(n,t'|i_0),
\end{eqnarray} 
This is the master equation (\ref{genreset}) for 
the single particle. In this paper, $i_0=0$ and we renote 
$P(n,t|i_0)\rightarrow P(n,t)$ for convenience. 

\section{Memory term in the subdiffusive case}\label{app_mem}

We choose a fixed $\epsilon\ll1$ and $t$ large enough
so that $\epsilon t\gg 1$. Hence, the memory term of 
equation (\ref{genM2b}) for the evolution of $M_2(t)$ reads:
\begin{eqnarray}
{\cal F}\{Kt^{\nu}\}&=& K\sum_{\tau=0}^{t}
\frac{(t-\tau)^{\nu}}{(1+\tau)^{\beta}}= 
Kt^{\nu}\sum_{\tau=0}^{t}\frac{(1-\tau/t)^{\nu}}{(1+\tau)^{\beta}}\nonumber\\
&=&Kt^{\nu}\left[\sum_{\tau=0}^{\epsilon t}\frac{(1-\tau/t)^{\nu}}
{(1+\tau)^{\beta}} +\sum_{\tau=\epsilon t+1}^{t}
\frac{(1-\tau/t)^{\nu}}{(1+\tau)^{\beta}} \right]\nonumber\\
&\simeq& Kt^{\nu}\left[\sum_{\tau=0}^{\epsilon t}\frac{1}{(1+\tau)^{\beta}}
-\frac{\nu}{t}\sum_{\tau=0}^{\epsilon t}\frac{\tau}{(1+\tau)^{\beta}}\right.
\nonumber\\
&+&\left.\sum_{\tau=\epsilon t+1}^{t}\frac{(1-\tau/t)^{\nu}}
{(1+\tau)^{\beta}}\label{F2a}\right].
\end{eqnarray}
The first sum in the brackets of (\ref{F2a}) can be written as
\begin{eqnarray}
\sum_{\tau=0}^{\epsilon t}\frac{1}{(1+\tau)^{\beta}}&=&\zeta(\beta)
-\sum_{\tau=\epsilon t+1}^{\infty}\frac{1}{(1+\tau)^{\beta}}\\
&\simeq& \zeta(\beta)-\int_{\epsilon t}^{\infty}d\tau(1+\tau)^{-\beta}\\
&\simeq& \zeta(\beta)-\frac{(\epsilon t)^{-\beta+1}}{\beta-1},\label{T1}
\end{eqnarray}
owing to the Euler-Maclaurin expansion of the discrete sum. 
Similarly, the third sum in 
the brackets of (\ref{F2a}) reads:
\begin{eqnarray}
\sum_{\tau=\epsilon t+1}^{t}\frac{(1-\tau/t)^{\nu}}{(1+\tau)^{\beta}}
&\simeq&\int_{\epsilon t}^{t}d\tau\frac{(1-\tau/t)^{\nu}}{(1+\tau)^{\beta}}\\
&\simeq&t^{-\beta+1}\int_{\epsilon}^1du\frac{(1-u)^{\nu}}{u^{\beta}}\label{T3}\\
&=&O((\epsilon t)^{-\beta+1}).
\end{eqnarray}
The second term in the brackets of (\ref{F2a}) is:
\begin{equation}
\frac{\nu}{t}\sum_{\tau=0}^{\epsilon t}\frac{\tau}{(1+\tau)^{\beta}}
=O\left(t^{-1}\int^{\epsilon t}d\tau \tau^{1-\beta}\right)
=O\left(\epsilon (\epsilon t)^{-\beta+1}\right),
\end{equation}
and thus negligible compared to (\ref{T1}) and (\ref{T3}).
We then substitute (\ref{T1}) and (\ref{T3}) in (\ref{F2a}), and
Eq. (\ref{F2}) is obtained after taking the limit $\epsilon\rightarrow0$.



\begin{thebibliography}{40}

\bibitem{annals}
E. Bolthausen and U. Schmock, Ann. Probab. {\bf 25}, 531 (1997).

\bibitem{siam}
H. G. Othmer and A. Stevens,
SIAM J. Appl. Math. {\bf 57}, 1044 (1997).

\bibitem{pemantle}
R. Pemantle, Probab. Surv. {\bf 4} 1, (2007).

\bibitem{gautestad2005}
A. O. Gautestad and I. Mysterud, Am. Nat. {\bf 165}, 44 (2005).

\bibitem{gautestad2006}
A. O. Gautestad and I. Mysterud, Ecol. Complex. {\bf 3}, 44 (2006).

\bibitem{davis}
B. Davis, Probab. Theor. Related Fields {\bf 84}, 203 (1990).

\bibitem{havlin}
A. Ordemann, E. Tomer, G. Berkolaiko, S. Havlin, and A. Bunde, 
Phys. Rev. E {\bf 64}, 046117 (2001).

\bibitem{woo}
J. W. Lee, J. Phys A: Math. Gen. {\bf 31}, 3929 (1998). 

\bibitem{grassberger}
J. G. Foster, P. Grassberger, and M. Paczuski, 
New J. Phys. {\bf 11}, 023009 (2009).

\bibitem{epl}
J. Choi, J. I. Sohn, K. I. Goh, I. M. Kim,
EPL {\bf 99}, 50001 (2012).

\bibitem{bouchaud}
J.-P. Bouchaud and A. Georges, Phys. Rep. {\bf 195}, 127 (1990).

\bibitem{klafter}
R. Metzler and J. Klafter, Phys. Rep. {\bf 339}, 1 (2000).

\bibitem{elephant1}
G. M. Sch${\rm \ddot{u}}$tz and S. Trimper,
Phys. Rev. E {\bf 70}, 045101 (R) (2004).

\bibitem{elephant2}
J. C. Cressoni, M. A. A. da Silva, and G. M. Viswanathan,
Phys. Rev. Lett. {\bf 98}, 070603 (2007).

\bibitem{italian}
M. Serva, Phys. Rev. E {\bf 88}, 052141 (2013).

\bibitem{peliti}
L. Peliti, J. Phys {\bf 46}, 1469 (1985).

\bibitem{nuovo}
L. Peliti and L. Pietronero, Riv. Nuovo Cimento {\bf 10}, 1 (1987).

\bibitem{quasistatic}
V. B. Sapozhdcov, J. Phys. A: Math. Gen. {\bf 27}, L151 (1994).

\bibitem{song} 
C. Song, T. Koren, P. Wang and A.-L. Barab\'asi, 
Nature Phys. {\bf 6}, 818 (2010).

\bibitem{nathan}
R. Nathan {\it et al.},
Proc. Natl. Acad. Sci. USA {\bf 105}, 19052 (2008).

\bibitem{sims}
D. W. Sims {\it et al.}, Nature {\bf 451}, 1098 (2008).

\bibitem{weimer}
N. E. Humphries, H. Weimerskirch, N. Queiroz, E. J. Southall, 
and D. W. Sims, Proc. Natl. Acad. Sci. USA {\bf 109}, 7169 (2012).

\bibitem{geisel}
D. Brockmann, L. Hufnagel and T. Geisel, 
Nature {\bf 439}, 462 (2006).

\bibitem{gonzalez}
M. C. Gonz\'alez, C. A. Hidalgo and A.-L. Barab\'asi, 
Nature {\bf 453}, 779 (2008).

\bibitem{song1}
C. Song, Z. Qu, N. Blumm and A.-L. Barab\'asi,
Science {\bf 327}, 1018 (2010).

\bibitem{boyersolis}
D. Boyer and C. Solis-Salas,
Phys. Rev. Lett. {\bf 112}, 240601 (2014).

\bibitem{morales2014}
J. A. Merkle, D. Fortin, and J. M. Morales,
Ecol. Lett. {\bf 17}, 924 (2014).

\bibitem{moorcroftlewis} 
P. R. Moorcroft and M. A. Lewis, {\it Mechanistic home range analysis} 
(Princeton University Press, Princeton, 2006).

\bibitem{borger}
L. B${\rm\ddot{o}}$rger, B. D. Dalziel and J. M. Fryxell, 
Ecol. Lett. {\bf 11}, 637 (2008). 

\bibitem{vanmoorter}
B. van Moorter {\it et al.}, Oikos {\bf 118}, 641 (2009). 

\bibitem{boyerwalsh}
D. Boyer and P. D. Walsh, 
Phil. Trans. R. Soc. A {\bf 368}, 5645 (2010). 

\bibitem{oikos2013}
J. Nabe-Nielsen, J. Tougaard, J. Teilmann, K. Lucke, and M. C. Forchhammer,
Oikos {\bf 122}, 1307 (2013).

\bibitem{turchin}
P. Turchin, {\it Quantitative analysis of movement}.
(Sunderland, MA. Sinauer Associates Inc, 1998).

\bibitem{colding}
E. A. Codling, M. J. Plank, and S. Benhamou,
J. R. Soc. Interface {\bf 5}, 813 (2008).

\bibitem{randomsearch}
G. M. Viswanathan, M. G. E. da Luz, E. P. Raposo, and H. E. Stanley,
{\it The Physics of foraging} (Cambridge, Cambridge, 2011).

\bibitem{interm1}
O. B\'enichou, M. Coppey, M. Moreau, P.-H. Suet, and R. Voituriez,
Phys. Rev. Lett. {\bf 94}, 198101 (2005).

\bibitem{interm2}
O. B\'enichou, C. Loverdo, M. Moreau, and R. Voituriez,
Rev. Mod. Phys. {\bf 83}, 81 (2011).

\bibitem{interm3}
F. Bartumeus, Oikos {\bf 118}, 488 (2009).

\bibitem{interm4}
J. M. Morales, D. T. Haydon, J. Frair, K. E. Holsinger, J. M. Fryxell,
Ecology {\bf 85}, 2436 (2004).

\bibitem{levy1}
G. M. Viswanathan, E. P. Raposo, M. G. E. da Luz,
Phys. Life Rev. {\bf 5}, 133 (2008).

\bibitem{levy2}
F. Bartumeus, M. G. E. da Luz, G. M. Viswanathan, J. Catalan,
Ecology {\bf 86}, 3078 (2005).

\bibitem{rhee}
I. Rhee, M. Shin, S. Hong, K. Lee, and S. Chong, 
IEEE/ACM Trans. Netw. {\bf 19}, 630 (2011).

\bibitem{montroll}
E. W. Montroll and G. H. Weiss, J. Math. Phys. {\bf 6}, 167 (1965).
                                                                        
\bibitem{effecFP}
S. C. Lim and S. V. Muniandy, Phys. Rev. E {\bf 66}, 021114 (2002).

\bibitem{yule} 
G. U. Yule, Phil. Trans. R. Soc. (London) B {\bf 213}, 21 (1925)..

\bibitem{barabasi}
A.-L. Barab\'asi and R. Albert, Science {\bf 286}, 509 (1999).

\bibitem{leyvraz}
P. L. Krapivsky, S. Redner, and F. Leyvraz, Phys. Rev. Lett.
{\bf 85}, 4629 (2000).  

\bibitem{evansmaj}
M. R. Evans and S. N. Majumdar,
Phys. Rev. Lett. {\bf 106}, 160601 (2011).

\bibitem{gamma}
Namely, $B(x,y)=\Gamma(x)\Gamma(y)/\Gamma(x+y)$, $x\Gamma(x)=\Gamma(x+1)$
and $\Gamma(x)\Gamma(1-x)=\pi/\sin(\pi x)$.

\bibitem{shlesinger}
M. F. Shlesinger, J. Stat. Phys. {\bf 10}, 421 (1974).

\bibitem{kotulski}
M. Kotulski, J. Stat. Phys. {\bf 81}, 777 (1995).

\bibitem{he}
Y. He, S. Burov, R. Metzler, and E. Barkai, Phys. Rev. Lett. {\bf 101}, 
058101 (2008).

\bibitem{barkai}
A. Rebenshtok and E. Barkai, Phys. Rev. Lett. {\bf 99}, 210601 (2007).

\bibitem{sokolov}
A. Lubelski, I. M.  Sokolov,  and J. Klafter, 
Phys. Rev. Lett. {\bf 100}, 250602 (2008).

\bibitem{theil2014}
F. Thiel and I. M. Sokolov, Phys. Rev. E {\bf 89}, 012115 (2014).

\bibitem{jeong2014}
J.-H. Jeon, A. V.Checkin, R. Metzler,
arXiv:1405.2193 [cond-mat.stat-mech] (2014).

\bibitem{gandhiPRE}
M. A. A. da Silva, J. C. Cressoni, G. M. Sch${\rm \ddot{u}}$tz, 
G. M. Viswanathan, and S. Trimper,
Phys. Rev. E {\bf 88}, 022115 (2013).
 
\bibitem{interface}
D. Boyer, M. C. Crofoot and P. D. Walsh,
{\it J. R. Soc. Interface} {\bf 9}, 842-847 (2012).



%
%
%
%
%
%
%


\end{thebibliography}
\end{document}